\documentstyle[12pt]{article}

\begin{document}

\renewcommand{\theequation}{\thesection.\arabic{equation}}

\hsize37truepc\vsize61truepc
\hoffset=-.5truein\voffset=-0.8truein
\setlength{\baselineskip}{17pt plus 1pt minus 1pt}
\setlength{\textheight}{22.5cm}

\def\diag{{\rm diag}}
\def\I{{\rm i}}
\def\tr{{\rm tr}}
\def\boldsp{\mbox{\boldmath $\omega$}^A}
\def\boldalpha{\mbox{\boldmath $\alpha$}}
\def\boldbeta{\mbox{\boldmath $\beta$}}
\def\smallboldalpha{\mbox{\small \boldmath $\alpha$}}
\def\smallboldbeta{\mbox{\small \boldmath $\beta$}}
\def\boldomega{\mbox{\boldmath $\omega$}}
\def\rlx{\relax\leavevmode}
\def\inbar{\vrule height1.5ex width.4pt depth0pt}
\def\IC{\rlx\hbox{\,$\inbar\kern-.3em{\rm C}$}}
\def\smallfrac#1#2{\mbox{\small $\frac{#1}{#2}$}}

\input epsf

\begin{titlepage}

\noindent
July, 1994 \hfill{MRR042-94}\\
\mbox{ } \hfill{hep-th/9410042}
\vskip 1.6in
\begin{center}
{\Large {\bf Integrable vertex and loop models on the square}}\\[8pt]
{\Large {\bf lattice with open boundaries via reflection matrices}}
\end{center}

\normalsize
\vskip .3in

\begin{center}
C. M. Yung  \hspace{3pt}
and \hspace{3pt} M. T. Batchelor
\par \vskip .1in \noindent
{\it Department of Mathematics, School of Mathematical Sciences}\\
{\it Australian National University, Canberra ACT 0200, Australia}
\end{center}
%\footnotetext{ }
\par \vskip .25in

\begin{center}
{\Large {\bf Abstract}}\\
\end{center}
The procedure for obtaining integrable vertex models via reflection matrices
on the square lattice with open boundaries is reviewed and
explicitly carried out for a number of two- and three-state
vertex models. These models include the six-vertex model,
the 15-vertex $A_2^{(1)}$ model and the 19-vertex models
of Izergin-Korepin and Zamolodchikov-Fateev. In each case the eigenspectra
is determined by application of either the algebraic or the analytic Bethe
ansatz with inhomogeneities. With suitable choices of reflection matrices,
these vertex models can be associated with integrable loop models on the
same lattice. In
general, the required choices {\em do not} coincide with those which lead to
quantum group-invariant spin chains. The exact solution of the integrable
loop models -- including an $O(n)$ model on the square lattice with open
boundaries  -- is of relevance to the surface critical behaviour of
two-dimensional polymers.

\vspace{1cm}
%\begin{center}
{\bf Physics and Astronomy Classifications}. 05.50.+q
%\end{center}

\end{titlepage}

\section{Introduction}
\setcounter{equation}{0}

Vertex models are usually solved with periodic boundary conditions
\cite{Baxter82,Faddeev79,Kulish82,Gaudin83}. Periodic boundaries
ensure that the transfer matrix $\tau(u)$, formed as the
trace over the auxiliary space of the monodromy matrix $T(u)$,
forms a commuting family if $T(u)$ is constructed from solutions $R(u)$ of
the Yang-Baxter equation. The consequences of this commutativity are
well-known, and allows exact results to be obtained for various physical
quantities related to the models.

The coordinate Bethe ansatz --for vertex models-- (see e.g.\
Ref. \cite{Baxter82}), of course, pre-dates Baxter's
commuting transfer matrices. While simple enough for two-state models
like the six-vertex model, it becomes too cumbersome for models with a
higher number of states allowed on the edges.
This is to be contrasted with the quantum inverse scattering method
\cite{Faddeev79} and related algebraic and analytic Bethe ansatz techniques,
which rely on the commutativity of $\tau(u)$ in order to diagonalize it.
Nevertheless, the first case of a vertex model with non-periodic boundaries
was solved using the coordinate Bethe ansatz \cite{Owczarek89}. This was
for the six-vertex model on the lattice ${\cal L}$, being the square lattice
rotated by $45^{\circ}$ and with open boundaries on two sides
(Fig. 1).\footnote{We will sometimes refer to this lattice as the square
lattice with open boundaries.}
Subsequently \cite{Batchelor93} this has been extended to solve a
three-state vertex model on the same lattice.
The fact remains, however, that the method is somewhat unsystematic,
especially for open boundaries.

%%%%%%  Figure 1 here  %%%%%%%%%%%%%%%%%%%%%
\begin{figure}[htb]
\epsfxsize = 4cm
\vbox{\vskip .8cm\hbox{\centerline{\epsffile{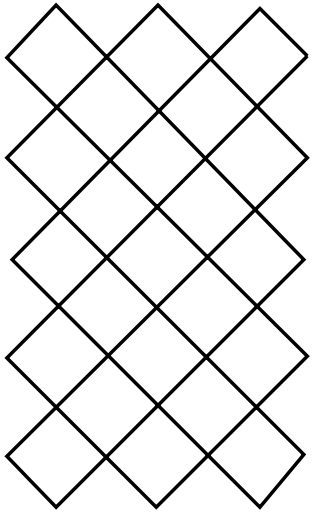}}}
%50%
\vskip .5cm \smallskip}
\caption{The lattice ${\cal L}$, of ``extent'' $M$ and $N$ ($M=10$ and
$N=6$ in the figure drawn) in the vertical and horizontal directions,
respectively. Boundary conditions are open in the horizontal direction
and periodic in the vertical.
}
%``occupied'' or ``empty'' edges.  }
\end{figure}
%%%%%%%%%%%%%%%%%%% End figure 1   %%%%%%%%%%%%%%%%%%%%%%%

On the other hand, Sklyanin \cite{Sklyanin88} has given a construction
of commuting transfer matrices $t(u,\boldomega)$, out of which he obtained the
Hamiltonian for the spin-$\smallfrac{1}{2}$ XXZ chain with open boundaries
(which has also been solved
using the coordinate Bethe ansatz \cite{Gaudin83,Alcaraz87}).
This transfer matrix $t(u,\boldomega)$ was constructed in the spirit of the
quantum inverse scattering method and involves the use of reflection
or $K$-matrices which satisfy a boundary version of the Yang-Baxter equation,
first discovered in the framework of factorized $S$-matrices
\cite{Cherednik84,Zamolodchikov}.
However $t(u,\boldomega)$ seemed a little mysterious as far as a vertex
model interpretation was concerned, and it was not until fairly recently
that it was shown \cite{Destri92}
that by specializing the inhomogeneities $\omega_i$
in a specific manner it becomes precisely the transfer matrix $t_D(u)$
for a vertex model on ${\cal L}$
(i.e.\ the ``diagonal-to-diagonal transfer matrix'' with open boundaries; see
Fig. 2). Therefore the Sklyanin scheme
``explains'' why the coordinate Bethe ansatz works for open boundaries in
the same way that Baxter's
 commuting transfer matrices $\tau(u)$ ``explains'' why
it works for periodic boundaries.

%%%%%%%%%%%%%%%%%%  Figure 2 here  %%%%%%%%%%%%%%%%%%%%%
\begin{figure}[htb]
\epsfxsize = 7cm
\vbox{\vskip .8cm\hbox{\centerline{\epsffile{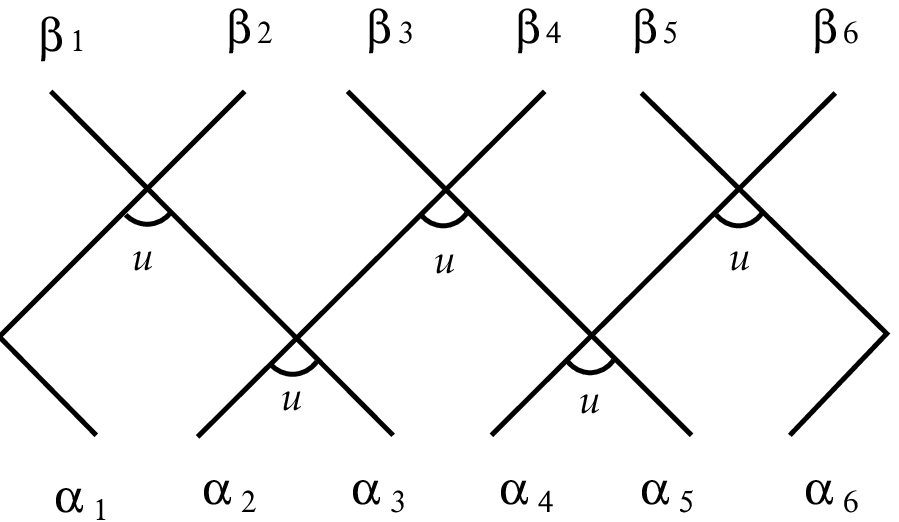}}}
\vskip .5cm \smallskip}
\caption{The ``diagonal-to-diagonal''
transfer matrix $t_D(u)$ which takes
the state $(\alpha_1,\ldots,\alpha_N)$ to the state
$(\beta_1,\ldots,\beta_N)$. All internal edges are summed over.
}
\end{figure}
%%%%%%%%%%%%%%%%%%% End figure 2   %%%%%%%%%%%%%%%%%%%%%%%

By now the Sklyanin scheme has been generalized
\cite{Mezincescu91a,Mezincescu91b,Mezincescu92a}
to handle most of the $R$-matrices constructed from (affine) non-exceptional
Lie algebras \cite{Jimbo86,Bazhanov87}.
For some cases \cite{Mezincescu90,Mezincescu91b,deVega94a,deVega94b} the
transfer matrix $t(u,\boldomega)$ has been diagonalized with modified
versions of the algebraic or analytic Bethe ansatz, the emphasis always
being on the corresponding quantum chain with open boundaries.
In this
paper we use the Sklyanin scheme to obtain the corresponding integrable
vertex models on the lattice ${\cal L}$ and their Bethe ansatz solutions.
We will concentrate on several two- and three-state models, both for
simplicity and also because these appear to be the most interesting
physically. In particular we have in mind an application of these solutions
to loop models \cite{Nienhuis90} on the lattice ${\cal L}$, especially
those which have a bearing on the configurational statistics of polymers.

To be precise, the vertex models concerned will turn out to have
bulk weights $w_i$, $i=1,\ldots,19$ given in Fig. 3 and boundary weights
$w_i^L$ and $w_i^R$, $i=1,2,3$ given in Fig. 4.\footnote{
For a two-state model,
$w_i$, $i=1,\ldots,13$, $w_2^R$ and $w_2^L$ are necessarily zero.}
The vertex model is defined through the partition function
\begin{equation}
  Z_{\rm vertex} = \sum_{\rm configs} \; \prod_{\rm vertices}
     \left({\rm Boltzmann \; weights}\right),
\label{eqn:vp}
\end{equation}
in the usual way. With periodic boundary conditions in the vertical
direction, this partition function is equivalent to the trace of a
product of transfer matrices $t_D(u)$, which then becomes the central
object to study.

The outline of the paper is as follows: In Section 2 we explain Sklyanin's
method in greater detail
and show how to obtain integrable vertex models on the lattice
${\cal L}$. In the following few sections we carry out the procedure explicitly
and obtain integrable open boundary versions of the six-vertex model,
the Zamolodchikov-Fateev 19-vertex model, the
$A_2^{(1)}$ model and the Izergin-Korepin
model, respectively. The corresponding Bethe ansatz solutions are also given.
In Section 7 we show how some of these models on ${\cal L}$ have an
interpretation as loop models on the same lattice. Bethe ansatz solutions
for these integrable loop models with open boundaries are given together
with their intended applications. Finally we conclude in Section 8 with a
discussion of several open problems related to vertex models on ${\cal L}$.

%%%%%%%%%%%%%%%%%%  Figure 3 here  %%%%%%%%%%%%%%%%%%%%%
\begin{figure}[htb]
\epsfxsize = 12cm
\vbox{\vskip .8cm\hbox{\centerline{\epsffile{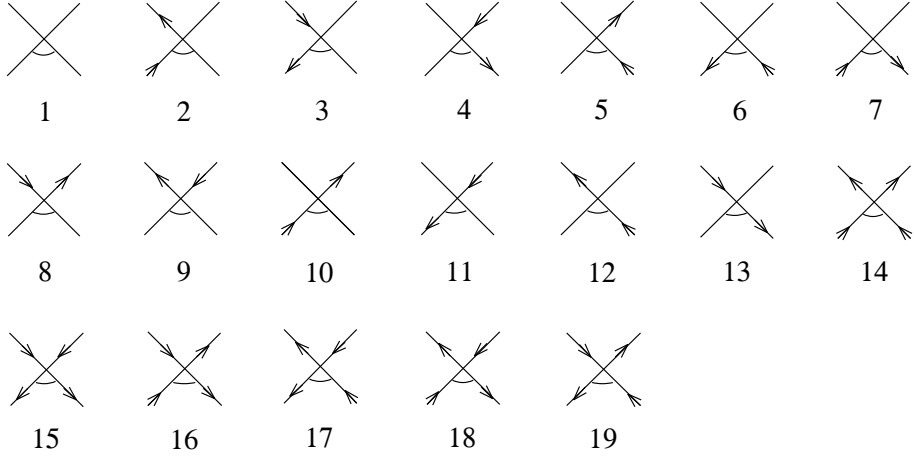}}}
%50%
\vskip .5cm \smallskip}
\caption{Allowed bulk vertices for integrable two- and three-state vertex
models studied in this paper. Vertex $i$ has an associated Boltzmann weight
$w_i$.}
\end{figure}
%%%%%%%%%%%%%%%%%%% End figure 3   %%%%%%%%%%%%%%%%%%%%%%%

%%%%%%%%%%%%%%%%%%  Figure 4 here  %%%%%%%%%%%%%%%%%%%%%
\begin{figure}[htb]
\epsfxsize = 8cm
\vbox{\vskip .8cm\hbox{\centerline{\epsffile{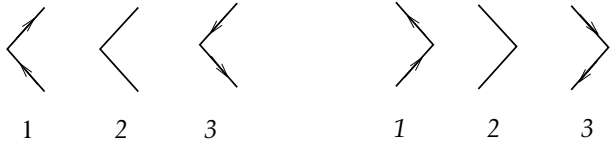}}}
%50%
\vskip .5cm \smallskip}
\caption{Allowed boundary vertices for the vertex models concerned.
The
left (respectively, right)
boundary vertices have Boltzmann
weights $w_i^L$ (respectively, $w_i^R$).
}
\end{figure}
%%%%%%%%%%%%%%%%%%% End figure 4   %%%%%%%%%%%%%%%%%%%%%%%

\section{$K$-matrices and the diagonal-to-diagonal transfer matrix}
\setcounter{equation}{0}

Let us first review the case of integrability in the bulk, i.e.\ for
vertex models on a standard square lattice with periodic boundary conditions.
Such an integrable $n$-state vertex model is specified by an $R$-matrix
$R(u)$ which acts on $\IC^n \otimes \IC^n$ and satisfies the Yang-Baxter
equation
\begin{equation}
R_{12}(u) R_{13}(u+v) R_{23}(v) = R_{23}(v) R_{13}(u+v) R_{12}(u),
\label{eqn:ybe}
\end{equation}
where $R_{12}(u)$, $R_{13}(u)$ and $R_{23}(u)$ act on $\IC^n \otimes
\IC^n \otimes \IC^n$, with $R_{12}(u)=R(u) \otimes 1$,
$R_{23}(u)=1\otimes R(u)$ etc. The vertex weights are determined from the
matrix
elements of $R(u)$ as shown in Fig. 5. For a two-state vertex model, the
states 1 and 2 will be associated with an
up-arrow and a down-arrow, respectively,
while for a three-state model, the states 1,2 and 3 will be associated
with an up-arrow, an empty edge and a down-arrow, respectively.
%%%%%%  Figure 5 here  %%%%%%%%%%%%%%%%%%%%%
\begin{figure}[htb]
\epsfxsize = 4.5cm
\vbox{\vskip .8cm\hbox{\centerline{\epsffile{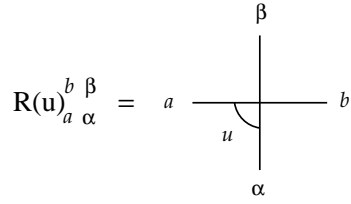}}}
%50%
\vskip .5cm \smallskip}
\caption{Graphical depiction of a matrix element of $R(u)$. The indices run
from 1 to $n$ for an $n$-state model.}
\end{figure}
%%%%%%%%%%%%%%%%%%% End figure 5   %%%%%%%%%%%%%%%%%%%%%%%

 Define the monodromy matrices
\begin{eqnarray}
 T(u,\boldomega) & = & R_{a1}(u+\omega_1)R_{a2}(u+\omega_2) \cdots R_{aN}(u+
   \omega_N), \label{eqn:monod1}\\
 \tilde{T}(u,\boldomega) & = & R_{Na}(u-\omega_N)\cdots R_{2a}(u-\omega_2)
  R_{1a}(u-\omega_1),
\label{eqn:monod2}
\end{eqnarray}
where $a$ denotes the auxiliary space and the quantities $\omega_i$ are
``inhomogeneities''. Graphically, $a$ labels the horizontal
direction whereas $1,\ldots,N$ label the columns of the square lattice
(from left to right).
Due to the Yang-Baxter equation (\ref{eqn:ybe}), the monodromy matrix
 $T(u,\boldomega)$ satisfies the intertwining relation
\begin{equation}
  R_{12}(u-v) \stackrel{1}{T}(u,\boldomega) \stackrel{2}{T}(v,\boldomega) =
 \stackrel{2}{T}(v,\boldomega) \stackrel{1}{T}(u,\boldomega)  R_{12}(u-v),
\end{equation}
where $\stackrel{1}{T}(u,\boldomega)= T(u,\boldomega) \otimes 1$ and
$\stackrel{2}{T}(u,\boldomega)= 1 \otimes T(u,\boldomega)$ etc. It then
follows that the row-to-row transfer matrix
$\tau(u,\boldomega)=\tr \; T(u,\boldomega)$,
where the trace is over the auxiliary space, forms a commuting family:
\begin{equation}
[\tau(u,\boldomega),\tau(v,\boldomega)]=0.
\end{equation}

To determine an integrable vertex model on the lattice ${\cal L}$, it is
first necessary that the bulk vertex weights be specified by an $R$-matrix
$R(u)$ as described above -- together
with a $+45^{\circ}$ rotation (cf.\ Fig. 3).
The boundary weights then follow, as we shall
see, from $K$-matrices which satisfy boundary versions of the Yang-Baxter
equation \cite{Sklyanin88,Destri92}.  We will require
\cite{Mezincescu91b} that $R(u)$ satisfies the following properties
\begin{eqnarray}
{\rm regularity} &:& R_{12}(0) = \rho(0)^{1/2} {\cal P}_{12},\\
{\rm unitarity} & : & R_{12}(u)\; R_{12}^{t_1 t_2}(-u) = \rho(u),
\label{eqn:unit}\\
PT-{\rm symmetry} & : & {\cal P}_{12} R_{12}(u) {\cal P}_{12} = R_{12}(u)^{t_1
  t_2},\\
{\rm crossing-symmetry} & :& R_{12}(u) = \stackrel{1}{V} R_{12}^{t_2}(-u-\eta)
  \stackrel{1}{V^{-1}}.
\end{eqnarray}
Here, $\rho(u)$ is an even scalar function,
${\cal P}$ is the exchange operator defined by
${\cal P}(f \otimes g) = (g \otimes f)$, $t_i$ denotes transposition in
the space $i$, $\eta$ is the crossing parameter and $V$ determines the
crossing matrix $M \equiv V^t V = M^t$. The crossing-symmetry can be replaced
by a weaker symmetry \cite{Mezincescu92a}, as will be necessary when we
consider the $A_2^{(1)}$ model in Section 5. We note that unitarity and
crossing-symmetry together imply the useful relation
\begin{equation}
    \stackrel{1}{M} R_{12}^{t_2}(-u-\eta) \stackrel{1}{M^{-1}}
  R_{12}^{t_1}(u-\eta) = \rho(u).
\label{eqn:m}
\end{equation}
Furthermore, we see that unitarity implies $T(u,\boldomega)\tilde{T}
 (-u,\boldomega) = \prod_{i=1}^N \rho(u-\omega_i)$. Therefore, up to a
scalar factor, $\tilde{T}(-u,\boldomega)$ is the
inverse of $T(u,\boldomega)$. Both monodromy matrices (\ref{eqn:monod1})
and (\ref{eqn:monod2}) are necessary in order to
construct integrable models on the lattice ${\cal L}$.

The boundary versions of the Yang-Baxter equation are given by
\cite{Sklyanin88,Mezincescu91b}
\begin{eqnarray}
\lefteqn{R_{12}(u-v) \stackrel{1}{K^-}(u) R_{12}^{t_1 t_2}(u+v)
  \stackrel{2}{K^-}(v)=}\hspace{50pt}\nonumber\\
& & \stackrel{2}{K^-}(v) R_{12}(u+v) \stackrel{1}{K^-}(u) R_{12}^{t_1
t_2}(u-v),
\label{eqn:k1}\\
\lefteqn{R_{12}(-u+v) \stackrel{1}{(K^+)^{t_1}}(u) \stackrel{1}{M^{-1}}
R_{12}^{t_1 t_2}(-u-v-2\eta) \stackrel{1}{M} \stackrel{2}{(K^+)^{t_2}}(v) = }
\hspace{50pt}\nonumber\\ & &
\stackrel{2}{(K^+)^{t_2}}(v) \stackrel{1}{M} R_{12}(-u-v-2\eta)
\stackrel{1}{M^{-1}}  \stackrel{1}{(K^+)^{t_1}}(u) R_{12}^{t_1 t_2}(-u+v),
\label{eqn:k2}
\end{eqnarray}
where $\stackrel{1}{K^-}(u)=K^-(u)\otimes 1$, $\stackrel{2}{K^-}(u)=1\otimes
K^-(u)$ etc. In practice, if $K^-(u)$ is a solution of (\ref{eqn:k1}) then
\begin{equation}
K^+(u)=K^-(-u-\eta)^t M
\label{eqn:auto}
\end{equation}
is a solution of (\ref{eqn:k2}).
Define the ``doubled monodromy matrix'' $U(u,\boldomega)$ by
\begin{equation}
U(u,\boldomega) = T(u,\boldomega) K^-(u) \tilde{T}(u,\boldomega),
\label{eqn:doub}
\end{equation}
with the multiplication in the auxiliary space $a$.
By (\ref{eqn:k1}) we have the ``intertwining relation''
\begin{eqnarray}
\lefteqn{R_{12}(u-v) \stackrel{1}{U}(u,\boldomega)
R_{12}^{t_1 t_2}(u+v)\stackrel{2}{U}(v,\boldomega)  = }\hspace{50pt}\nonumber\\
& & \stackrel{2}{U}(v,\boldomega) R_{12}(u+v)
\stackrel{1}{U}(u,\boldomega) R_{12}^{t_1 t_2}(u-v).
\label{eqn:int}
\end{eqnarray}

The main result is the following: If the boundary equations (\ref{eqn:k1})
and (\ref{eqn:k2}) are satisfied, then Sklyanin's transfer matrix
\cite{Sklyanin88,Mezincescu91b}
\begin{equation}
 t(u,\boldomega) = \tr\; K^+(u) U(u,\boldomega),
\label{eqn:skl}
\end{equation}
where the trace is again in the auxiliary space $a$,
forms a commuting family:
\begin{equation}
[t(u,\boldomega),t(v,\boldomega)]=0.
\end{equation}
The proof of commutativity \cite{Sklyanin88} involves showing that the object
\begin{equation}
\tr_{12} \;B_{12}(u,v)^{t_1 t_2} A_{12}(u,v),
\end{equation}
where $A_{12}(u,v)$ and $B_{12}(u,v)$ are the left (respectively, right)
hand sides of (\ref{eqn:k2}) and (\ref{eqn:int}), can be manipulated using
unitarity (\ref{eqn:unit}), equation (\ref{eqn:m}), and the properties of
the trace $\tr$ and transposition $t_i$
 to obtain $t(u,\boldomega)t(v,\boldomega)$
(respectively, $t(v,\boldomega)t(u,\boldomega)$) times the scalar factor
$\rho(u-v)\rho(u+v+\eta)$.
The transfer matrix $t(u,\boldomega)$ has an interesting graphical
interpretation \cite{Destri92}. Let $\boldalpha=(\alpha_1,\ldots,\alpha_N)$
denote a state in ($\IC^n)^{\otimes N}$. Then the matrix element
$t(u,\boldomega)^{\smallboldbeta}_{\smallboldalpha}$ is given by
\begin{equation}
t(u,\boldomega)_{\smallboldalpha}^{\smallboldbeta} = K^+_{ab}(u)
\left(T_{bc}(u,\boldomega)\right)_{\alpha_1\ldots\alpha_N}
^{\gamma_1\ldots\gamma_N} K^-_{cd}(u)
\left(\tilde{T}_{da}(u,\boldomega)\right)_{\gamma_1\ldots\gamma_N}
^{\beta_1\ldots\beta_N},
\end{equation}
and is represented graphically by Fig. 6.

%%%%%%%%%%%%%%%%%%  Figure 6 here  %%%%%%%%%%%%%%%%%%%%%
\begin{figure}[htb]
\epsfxsize = 9cm
\vbox{\vskip .8cm\hbox{\centerline{\epsffile{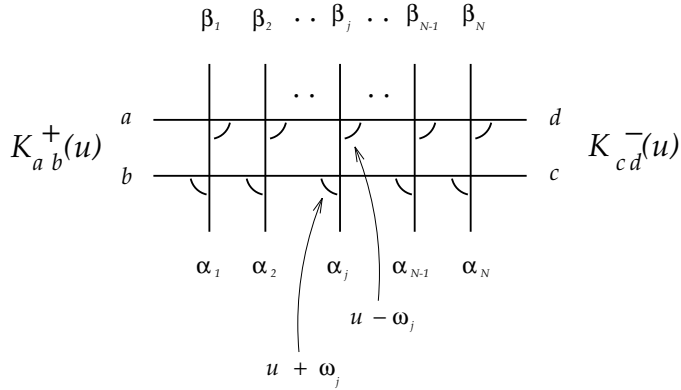}}}
\vskip .5cm \smallskip}
\caption{The Sklyanin transfer matrix $t(u,\omega)$.
The positioning of the ``rapidities'' $u\pm \omega_i$ is crucial.
All internal edges (including the ones labelled by $a,b,c,d$ are
summed over.}
\end{figure}
%%%%%%%%%%%%%%%%%%% End figure 6   %%%%%%%%%%%%%%%%%%%%%%%

The commutativity of $t(u,\boldomega)$ implies integrability of the
open quantum spin chain with Hamiltonian \cite{Sklyanin88}
\begin{equation}
 H = \sum_{k=1}^{N-1} H_{k,k+1} + \smallfrac{1}{2} \stackrel{1}{K^t_-}(0)
  + \frac{\tr_0 \stackrel{0}{K_+}(0) H_{N,0}}{\tr K_+(0)}
\label{eqn:ham}
\end{equation}
whose two-site terms are given by
\begin{equation}
H_{k,k+1} = \left. \frac{d}{du} {\cal P}_{k,k+1} R_{k,k+1}(u)\right|_{u=0},
\end{equation}
in the standard fashion.\footnote{If the $R$ matrix is in the ``homogeneous
gauge'' so that $[\check{R}_{12}(u),\check{R}_{12}(v)]=0$, then the choice
$K^-(u)=1$, $K^+(u)=M$ yields a quantum group-invariant Hamiltonian
\cite{Mezincescu91c,Pasquier90}.}
However it is not clear at this stage in what sense
$t(u,\boldomega)$ is a ``transfer matrix''. One answer is the following
\cite{Destri92}: If the inhomogeneities $\omega_i$ are chosen appropriately
then the transfer matrix can be interpreted as that for a vertex model on the
lattice ${\cal L}$. More specifically, define the function $\boldsp(u)$ by
\begin{equation}
\left(\boldsp(u)\right)_j = (-)^{j+1} u,
\label{eqn:alt}
\end{equation}
$j=1,\ldots,N$.
Then $t(u,\boldsp(u))$ is related to the transfer matrix $t_D(u)$
for a vertex model on the lattice ${\cal L}$. To see this it is most
instructive
to proceed graphically. Firstly, note that regularity of $R(u)$ implies that
$R(0)$ has the graphical description given in Fig. 7.
%%%%%%%%%%%%%%%%%%  Figure 7 here  %%%%%%%%%%%%%%%%%%%%%
\begin{figure}[htb]
\epsfxsize = 5.5cm
\vbox{\vskip .8cm\hbox{\centerline{\epsffile{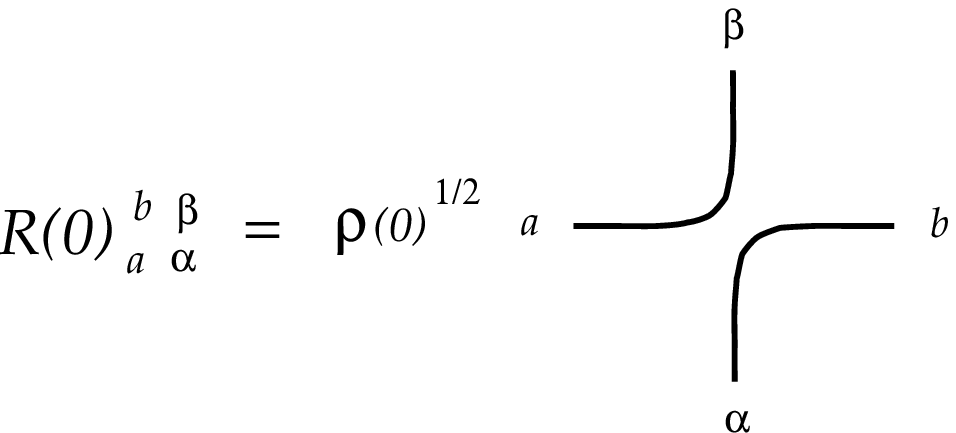}}}
\vskip .5cm \smallskip}
\caption{Graphical depiction of $R(0)_{a \; \alpha}^{b\;\beta} = \rho(0)^{1/2}
  \delta_a^{\beta}\delta_{\alpha}^{b}$.}
\end{figure}
%%%%%%%%%%%%%%%%%%% End figure 7   %%%%%%%%%%%%%%%%%%%%%%%
It is then apparent from Fig. 6 that
$t(u,\boldsp(u))_{\boldalpha}^{\boldbeta}$ has the graphical interpretation
given in Fig. 8 (assuming $N$ is even).
Define now the ``left and right boundary weights''
\begin{eqnarray}
  l_a^b(u) & = & \sum_{cd} R_{da}^{bc}(2u) K^+_{cd}(u),\nonumber\\
  r_a^b(u) & = & K^-_{ab}(u),
\label{eqn:lr}
\end{eqnarray}
which are depicted graphically in Fig. 9.
%%%%%%%%%%%%%%%%%%  Figure 8 here  %%%%%%%%%%%%%%%%%%%%%
\begin{figure}[htb]
\epsfxsize = 11cm
\vbox{\vskip .8cm\hbox{\centerline{\epsffile{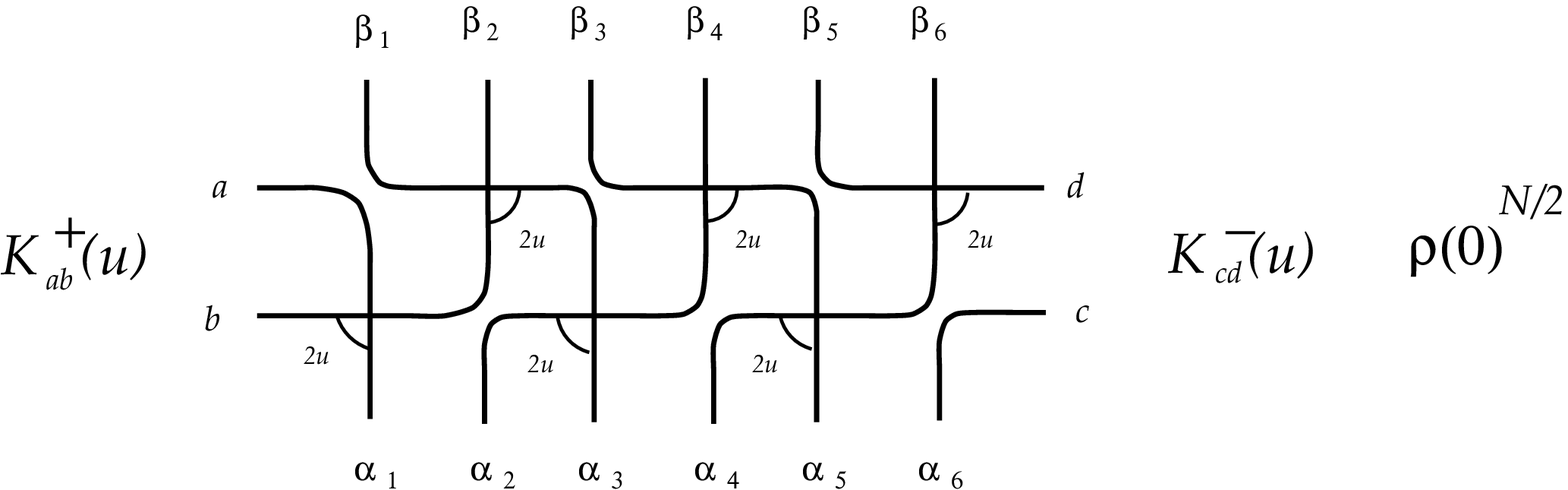}}}
%50%
\vskip .5cm \smallskip}
\caption{Graphical depiction of $t(u,\omega^A(u))$, the Sklyanin transfer
matrix with alternating inhomogeneities. $N=6$ in this figure.}
\end{figure}
%%%%%%%%%%%%%%%%%%% End figure 8   %%%%%%%%%%%%%%%%%%%%%%%
It is now clear that
Fig. 8  can be deformed into Fig. 2 (modulo an overall scalar
factor $\rho(0)^{N/2}$), upon replacing $u$ by  $u/2$. The
naming of $l^b_a(u)$ and $r_a^b(u)$ is also justified even though, strictly
speaking, these are boundary weights for a model with ``spectral parameter''
$2u$. The
left and right boundary weights can be separately normalized, as can the
bulk weights determined by the $R$-matrix. In practice, we will find it
convenient to choose a normalization such that the bulk weight $w_{14}$ and
boundary weights $w_1^R$ and $w_1^L$ (i.e.\ where all edges are in the up-arrow
state) are all unity. Thus we have the key relationship
\begin{equation}
t_D(u) = \rho(0)^{-N/2}\left(R_{11}^{11}(u)\right)^{-N+1}
  \frac{t\left(u/2,\boldsp(u/2)\right)}{l_1^1(u/2)
   r_1^1(u/2)}
\label{eqn:key}
\end{equation}
which defines an integrable vertex model on ${\cal L}$. The eigenvalues
$\Lambda_D(u)$ of this transfer matrix are obtained once we have diagonalized
the Sklyanin transfer matrix $t(u,\boldomega)$. To do this, the algebraic
or the analytic Bethe ansatz (suitably generalized)
will need to be used, depending on the model.

Given an $R$-matrix $R(u)$ which satisfies the Yang-Baxter equation, there
are typically many inequivalent solutions to (\ref{eqn:k1}) for $K^-(u)$
(which are, of course, defined only up to an overall scalar factor since
(\ref{eqn:k1}) is a homogeneous equation). In this paper we will only be
concerned with diagonal $K$-matrices. These lead to boundary weights
$r_a^b(u)$ and $l_a^b(u)$ which are non-zero only if $a=b$; i.e.\ to models
for which there is ``conservation of arrows across diagonals'' and which can,
in principle (but hardly in practice) also be solved with the coordinate
Bethe ansatz.

%%%%%%%%%%%%%%%%%%  Figure 9 here  %%%%%%%%%%%%%%%%%%%%%
\begin{figure}[htb]
\epsfxsize = 8cm
\vbox{\vskip .8cm\hbox{\centerline{\epsffile{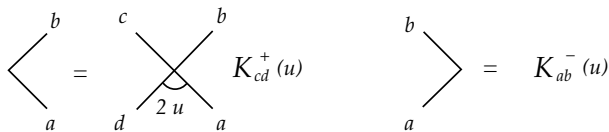}}}
%50%
\vskip .5cm \smallskip}
\caption{Graphical depiction  of left and right boundary weights
$l_a^b(u)$ and $r_a^b(u)$.}
\end{figure}
%%%%%%%%%%%%%%%%%%% End figure 9   %%%%%%%%%%%%%%%%%%%%%%%

The ``alternating inhomogeneities'' $\boldsp(u)$ also play a similar role
in constructing diagonal-to-diagonal transfer matrices from inhomogeneous
row-to-row ones in the case of periodic boundary conditions (cf.\ Refs.
\cite{Truong83,Destri89}). It is important to appreciate that we {\em do not}
have $[t_D(u),t_D(v)]=0$ for the chosen $u$-dependent inhomogeneities.
Thus we see that ``integrability'' on the lattice ${\cal L}$ is inherited from
a more abstract problem -- commutativity of Sklyanin transfer matrices --
instead of being manifestly so.

\section{The six-vertex model}
\setcounter{equation}{0}

In Section 2 we described a procedure based on the quantum inverse scattering
method for obtaining an integrable vertex model on the square
lattice with open boundaries, related to well-known models integrable on the
square lattice with {\em periodic} boundaries.
This procedure has not previously been carried out in
its entirety for any vertex model, although it has been pushed through
\cite{Destri92} for the six-vertex model in a special case (with one special
choice of $K$-matrices). In fact, solving this model with a general $K$-matrix
does not involve much more work, especially
since the diagonalization of $t(u,\boldomega)$ has already been done in Ref.
\cite{Sklyanin88}. We include it here for completeness and for pedagogical
reasons.

The six-vertex model is associated with the vector (spin-$\smallfrac{1}{2}$)
representation of the (quantum) affine Lie algebra $A_1^{(1)}$. Its
$R$-matrix reads:
\begin{equation}
R(u)=\left(\begin{array}{cccc}
    a(u) & & &  \\
    & b(u) & c(u) & \\
    & c(u) & b(u) & \\
    & & & a(u)
\end{array} \right),
\end{equation}
where
\begin{eqnarray}
a(u) & = & \sinh(u+\lambda), \nonumber\\
b(u) & = & \sinh(u), \nonumber\\
c(u) & = & \sinh(\lambda),
\end{eqnarray}
and from this the monodromy matrices $T(u,\boldomega)$ and $\tilde{T}
(u,\boldomega)$ are constructed according to (\ref{eqn:monod1}) and
(\ref{eqn:monod2}).
This $R$-matrix  has the properties of regularity and unitarity, with $\rho(u)=
\sinh(\lambda+u)\sinh(\lambda-u)$, $P$- and $T$-symmetry separately and
crossing-symmetry with $M=1$ and
$\eta=\lambda$. The most general diagonal solution
\cite{Cherednik84,Sklyanin88} for $K^-(u)$ is given (up to an unimportant
overall scalar function of $u$) by $K^-(u)=K(u,\xi_-)$ where
\begin{equation}
K(u,\xi) = \left( \begin{array}{c}
   \sinh(\xi+u) \hspace{60pt} \\
   \hspace{60pt} \sinh(\xi - u) \end{array} \right),
\label{eqn:km6}
\end{equation}
with $\xi$ being a free parameter.
Because of the automorphism (\ref{eqn:auto}), solutions for $K^+(u)$ are
given by $K^+(u)=K(-u-\lambda,\xi_+)$.
The transfer matrix (\ref{eqn:skl})  with these $K$-matrices
 has been diagonalized with the algebraic Bethe ansatz, modified for
open boundaries \cite{Sklyanin88}. A main step in this method
involves writing the doubled monodromy matrix $U(u,\boldomega)$ in the form
\begin{equation}
U(u,\boldomega) = \left( \begin{array}{cc}
   {\cal A}(u) & {\cal B}(u)\\
   {\cal C}(u) & {\cal D}(u)\end{array}\right)
\label{eqn:tbt}
\end{equation}
and using the ``intertwining relation'' (\ref{eqn:int})
to derive the commutation relations
\begin{eqnarray}
{\cal B}(u) {\cal B}(v) & = & {\cal B}(v) {\cal B}(u),\nonumber\\
{\cal A}(u) {\cal B}(v) &=& \frac{\sinh(\lambda+v-u) \sinh(u+v)}
  {\sinh(\lambda+u+v)\sinh(v-u)} {\cal B}(v){\cal A}(u) - \nonumber\\
&  & \frac{\sinh(u+v)\sinh(\lambda)}{\sinh(\lambda+u+v)\sinh(v-u)}
  {\cal B}(u){\cal A}(v) - \frac{\sinh(\lambda)}{\sinh(\lambda+u+v)}
{\cal B}(u) {\cal D}(v),\nonumber\\
\tilde{{\cal D}}(u){\cal B}(v) & = & - \frac{\sinh(u-v+\lambda)
  \sinh(u+v+2\lambda)}{\sinh(v-u)\sinh(\lambda+u+v)}{\cal B}(u)
  \tilde{{\cal D}}(u) \nonumber\\
& + & \frac{\sinh(\lambda)\sinh(2u+2\lambda)}
   {\sinh(v-u)\sinh(2v+\lambda)}{\cal B}(u) \tilde{{\cal D}}(v) \nonumber\\
& + & \frac{\sinh(\lambda)\sinh(2v)\sinh(2u+2\lambda)}
   {\sinh(2v+\lambda)\sinh(u+v+\lambda)}{\cal B}(u) {\cal A}(v),
\label{eqn:com}
\end{eqnarray}
where $\tilde{{\cal D}}(u) = \sinh(\lambda+2u) {\cal D}(u) - \sinh(\lambda)
 {\cal A}(u)$. The transfer matrix $t(u,\boldomega)$ can now be written
(using (\ref{eqn:skl}), (\ref{eqn:doub}) and (\ref{eqn:tbt})) as
\begin{equation}
t(u,\boldomega) = \frac{\sinh(\xi_+ + u + \lambda)}{\sinh(2u+\lambda)}
 \tilde{{\cal D}}(u) + \frac{\sinh(\xi_+ - u)\sinh(2u+2\lambda)}
 {\sinh(2u+\lambda)} {\cal A}(u).
\label{eqn:ad}
\end{equation}
The algebraic Bethe ansatz says that $t(u,\boldomega)$ is diagonal on the
states $| u_1,\ldots, u_M \rangle = {\cal B}(u_1) \ldots {\cal B}(u_M)
| \Omega \rangle$, where the pseudovacuum $| \Omega \rangle$ is the state
with all arrows up. The component ${\cal B}(u)$ of $U(u,\boldomega)$
acts as a creation operator
while ${\cal C}(u)$, which kills $| \Omega \rangle$, acts as an annihilation
operator. Furthermore both ${\cal A}(u)$ and ${\cal D}(u)$ are diagonal on
$| \Omega \rangle$. To obtain the eigenvalue $\Lambda(u,\boldomega)$ of
$t(u,\boldomega)$ on $| u_1,\ldots, u_M \rangle$, one uses the
commutation relations (\ref{eqn:com}) to move ${\cal A}(u)$ and
$\tilde{{\cal D}}(u)$ in (\ref{eqn:ad})
past the ${\cal B}(u_j)$, thus  obtaining the ``wanted terms''
\begin{eqnarray}
\Lambda(u,\boldomega) & = & \alpha(u,\boldomega)
  \frac{\sinh(\xi_+ - u)\sinh(2u+2\lambda)} {\sinh(2u+\lambda)}
  \prod_{j=1}^M \frac{\sinh(u - \lambda -u_j)\sinh(u+u_j)}
   {\sinh(u+\lambda+u_j)\sinh(u-u_j)}\nonumber\\
& + & \tilde{\delta}(u,\boldomega)
  \frac{\sinh(\xi_+ + u +\lambda)}{\sinh(2u+\lambda)}
\prod_{j=1}^M \frac{\sinh(u + \lambda -u_j)\sinh(u+2\lambda+u_j)}
   {\sinh(u+\lambda+u_j)\sinh(u-u_j)},
\label{eqn:eigxxz}
\end{eqnarray}
where $\alpha(u,\boldomega)$ and $\tilde{\delta}(u,\boldomega)$ are the
values taken on $| \Omega \rangle$ by ${\cal A}(u)$ and
$\tilde{{\cal D}}(u)$ respectively. The ``unwanted terms'' in this
procedure cancel (and $\Lambda(u,\boldomega)$ becomes analytic in $u$) only
if the numbers $u_k$ satisfy, for all $k$, the equations
\begin{eqnarray}
-\frac{\alpha(u_k,\boldomega)\sinh(u_k+\xi_+)\sinh(2u_k)}
  {\tilde{\delta}(u_k,\boldomega)\sinh(u_k - \xi_+ + \lambda)} & = &
\prod_{j \neq k}^M \frac{\sinh(u_k+u_j+2\lambda)
  \sinh(u_k-u_j + \lambda)}{\sinh(u_k+u_j)\sinh(u_k-u_j-\lambda)}.
\label{eqn:baexxz}
\end{eqnarray}
It is not hard to show - e.g.\ by graphical means (Fig. 6) - that
\begin{eqnarray}
\alpha(u,\boldomega) & = & \sinh(\xi_-+u) \prod_{i=1}^N a(u+\omega_i)
 a(u-\omega_i),\nonumber\\
\tilde{\delta}(u,\boldomega) & = & \sinh(\xi_- - u-\lambda)\sinh(2u)
\prod_{i=1}^N b(u+\omega_i)b(u-\omega_i).
\end{eqnarray}
After the shift $u_j \rightarrow u_j - \lambda/2$, the Bethe ansatz
equations (\ref{eqn:baexxz}) take on the symmetric form
\begin{eqnarray}
\lefteqn{
\left\{\frac{\sinh(u_k-\xi_+-\smallfrac{\lambda}{2})
  \sinh(u_k+\xi_- -\smallfrac{\lambda}{2})}
  {\sinh(u_k+\xi_+ +\smallfrac{\lambda}{2})
  \sinh(u_k- \xi_- +\smallfrac{\lambda}{2})}\right\}
\prod_{i=1}^N\frac{\sinh(u_k+\omega_i+\smallfrac{\lambda}{2})
   \sinh(u_k-\omega_i+\smallfrac{\lambda}{2})}
  {\sinh(u_k+\omega_i-\smallfrac{\lambda}{2})
   \sinh(u_k-\omega_i-\smallfrac{\lambda}{2})}}\hspace{170pt}\nonumber\\
& = & \prod_{j \neq k}^M\frac{\sinh(u_k+u_j+\lambda)\sinh(u_k-u_j+\lambda)}
   {\sinh(u_k+u_j-\lambda)\sinh(u_k-u_j-\lambda)}.
\label{eqn:sym}
\end{eqnarray}
The equations (\ref{eqn:eigxxz}) and (\ref{eqn:sym}) were obtained in
\cite{Sklyanin88} in the study of the spin-$\smallfrac{1}{2}$ XXZ chain with
open boundaries.

With the specialization of the inhomogeneities to $\boldomega=
\boldsp(u)$ defined by (\ref{eqn:alt}),
the second term of $\Lambda(u,\boldomega)$ in (\ref{eqn:eigxxz})
vanishes due to $\tilde{\delta}(u,\boldsp(u))=0$. Thus we have
(after the shift $u_j \rightarrow u_j - \lambda/2$)
\begin{equation}
\Lambda(u,\boldsp(u)) = \sinh(\xi_+ - u)\sinh(2u+2\lambda)\sinh(\xi_- + u)
  a(2u)^{N-1} \rho(0)^{N/2} A(u),
\label{eqn:eig6}
\end{equation}
where
\begin{equation}
A(u) = \prod_{j=1}^M \frac{\sinh(u-u_j-\smallfrac{\lambda}{2})
  \sinh(u+u_j-\smallfrac{\lambda}{2})}{\sinh(u+u_j+\smallfrac{\lambda}{2})
  \sinh(u-u_j+\smallfrac{\lambda}{2})}.
\label{eqn:au}
\end{equation}
{}From the definitions (\ref{eqn:lr})
we find that the left and right boundary weights are given by
\begin{eqnarray}
r_1^1(u) & = & \sinh(\xi_- + u) \nonumber\\
r_2^2(u) & = & \sinh(\xi_- - u) \nonumber\\
l_1^1(u) & = & \sinh(2u+2\lambda)\sinh(\xi_+ - u) \nonumber\\
l_2^2(u) & = & l_1^1(u) \frac{\sinh(\xi_+ +u)}{\sinh(\xi_+ - u)}.
\label{eqn:bo6v}
\end{eqnarray}
Therefore the eigenvalue $\Lambda_D(u)$ of the transfer matrix $t_D(u)$ is
(according to (\ref{eqn:key}) and (\ref{eqn:eig6}))
\begin{equation}
\Lambda_D(u) = A(u/2),
\end{equation}
with $A(u)$ defined by (\ref{eqn:au}). The associated Bethe ansatz equations
are given by
\begin{eqnarray}
\lefteqn{
\left\{\frac{\sinh(u_k-\xi_+-\smallfrac{\lambda}{2})
  \sinh(u_k+\xi_- -\smallfrac{\lambda}{2})}
  {\sinh(u_k+\xi_+ +\smallfrac{\lambda}{2})
  \sinh(u_k- \xi_- +\smallfrac{\lambda}{2})}\right\}
\left[\frac{\sinh(u_k+\smallfrac{u}{2}+\smallfrac{\lambda}{2})
   \sinh(u_k-\smallfrac{u}{2}+\smallfrac{\lambda}{2})}
  {\sinh(u_k+\smallfrac{u}{2}-\smallfrac{\lambda}{2})
   \sinh(u_k-\smallfrac{u}{2}-\smallfrac{\lambda}{2})}\right]^N}
\hspace{170pt}\nonumber\\
& = & \prod_{j \neq k}^M\frac{\sinh(u_k+u_j+\lambda)\sinh(u_k-u_j+\lambda)}
   {\sinh(u_k+u_j-\lambda)\sinh(u_k-u_j-\lambda)}.
\label{eqn:des}
\end{eqnarray}

When $\xi_{\pm} \rightarrow - \infty$, the term in braces on the left hand
side of (\ref{eqn:des}) becomes unity.\footnote{With this choice of
$K$-matrices, the associated open boundary
spin-$\smallfrac{1}{2}$ XXZ chain defined by (\ref{eqn:ham}) has
$U_q(su(2))$-invariance \cite{Pasquier90,Mezincescu91c,Destri92}.}
This is the case for which
an integrable vertex model on ${\cal L}$ has been previously worked out
\cite{Destri92} using the same procedure. This case has also been obtained
using
the coordinate Bethe ansatz in Ref. \cite{Owczarek89} and is referred to there
as the Potts model with free boundaries. In fact, the coordinate Bethe
ansatz also allows the general case to be solved; the Bethe ansatz equations
obtained in Ref. \cite{Owczarek89} can presumably be recast in the form
(\ref{eqn:des}).

\section{The Zamolodchikov-Fateev model}
\setcounter{equation}{0}
The simplest three-state vertex model is the Zamolodchikov-Fateev 19-vertex
\cite{Zamolodchikov80} or the
$A^{(1)}_1$ model in the symmetric rank-2 tensor (spin-$1$) representation
\cite{Kulish81} and can be constructed from the six-vertex model using the
fusion procedure. The associated quantum spin chain is the spin-$1$ XXZ
model. The $R$-matrix is given
by\footnote{This is also the $R$-matrix associated with the vector
representation of $B_1^{(1)}$ \cite{Jimbo86,Bazhanov87}.}
\begin{equation}
R(u)=\left(\begin{array}{ccc|ccc|ccc}
    a & & & & & & & & \\
    & b & & d & & & & &\\
    & & c & & e &  & f & &\\ \hline
    & d & & b & & & & &\\
    & & e & & g & & e & &\\
    & & & & & b & & d &\\ \hline
    & & f & & e & & c & &\\
    & & & & & d & & b &\\
    & & & & & & & & a \end{array} \right),
\end{equation}
with
\begin{eqnarray*}
a(u) & = & \sinh(u+2\lambda)\sinh(u+\lambda),\\
b(u) & = & \sinh(u)\sinh(u+\lambda),\\
c(u) & = & \sinh(u)\sinh(u-\lambda),\\
d(u) & = & \sinh(2\lambda)\sinh(u+\lambda),\\
e(u) & = & \sinh(u)\sinh(2\lambda),\\
f(u) & = & \sinh(\lambda)\sinh(2\lambda),\\
g(u) & = & b(u) + f(u).
\end{eqnarray*}
This $R$-matrix is regular and unitary, with $\rho(u)=\sinh(u+\lambda)
\sinh(u-\lambda)\sinh(u+2\lambda)\sinh(u-2\lambda)$,
$P$- and $T$-symmetric and crossing-symmetric with $M=1$ and $\eta=\lambda$.
The most general diagonal solution for $K^-(u)$ has been obtained in Ref.
\cite{Mezincescu90} and is given by $K^-(u)=K(u,\xi_-)$ where
\begin{equation}
K(u,\xi)=\left(\begin{array}{c}
    \sinh(u+\xi)\sinh(u-\lambda+\xi) \hspace{80pt}\\
    \hspace{10pt} -\sinh(u-\xi)\sinh(u-\lambda+\xi) \hspace{10pt}\\
    \hspace{80pt} \sinh(u-\xi)\sinh(u+\lambda-\xi)\end{array}\right).
\end{equation}
Once again, due to the automorphism (\ref{eqn:auto}) solutions for $K^+(u)$
are given by $K^+(u)=K^-(-u-\lambda,\xi_+)$.

The transfer matrix $t(u,\boldomega)$ with these general $K$-matrices but
 without inhomogeneities $\omega_j$ has been diagonalized in Ref.
\cite{Mezincescu90} by generalizing the approach used to solve the
corresponding model with periodic boundaries. The model with
inhomogeneities can be treated likewise. This involves the construction
of an auxiliary transfer matrix
\begin{equation}
{}_{\sigma}t(u,\boldomega)=\tr\left( {}_{\sigma}K^+(u) {}_{\sigma}T(u)
   {}_{\sigma}K^-(u) {}_{\sigma}\tilde{T}(-u)\right).
\end{equation}
which commutes with $t(u,\boldomega)$.
The auxiliary transfer matrix corresponds to that for a model
whose monodromy matrices are constructed
from  a simpler $R$-matrix ${}_{\sigma}R(u)$
and whose $K$-matrices ${}_{\sigma}K^{\pm}(u)$ are those from the
six-vertex model (\ref{eqn:km6}) , but with
the free parameters shifted according to $\xi_- \rightarrow \xi_- -
\smallfrac{\lambda}{2}$ and  $\xi_+ \rightarrow -\xi_+ +
\smallfrac{\lambda}{2}$. The $R$-matrix ${}_{\sigma}R(u)$ in turn is given
by
\begin{equation}
{}_{\sigma}R(u)=\left(\begin{array}{ccc|ccc}
    \bar{a}  & & & & & \\
    & \bar{b} & & \bar{d}  & &\\
    & & \bar{c} & & \bar{d} &  \\ \hline
    & & & & & \\[-10pt]
    & \bar{d} & & \bar{c} & & \\
    & & \bar{d} & & \bar{b} & \\
    & & & & & \bar{a}
\end{array} \right),
\end{equation}
with
\begin{eqnarray*}
\bar{a} & = & \sinh(u + \smallfrac{3\lambda}{2}),\\
\bar{b} & = & \sinh(u + \smallfrac{\lambda}{2}),\\
\bar{c} & = & \sinh(u - \smallfrac{\lambda}{2}),\\
\bar{d} & = & \left[\sinh(\lambda)\sinh(2\lambda)\right]^{1/2},
\end{eqnarray*}
and is associated to the tensor product of a spin-$\smallfrac{1}{2}$ and a
spin-$1$ representation of $A_1^{(1)}$ \cite{Kulish83}.

The monodromy matrices ${}_{\sigma}T(u)$ are $2 \times 2$ and thus
the auxiliary transfer matrix ${}_{\sigma}t(u,\boldomega)$
can be diagonalized with the algebraic Bethe ansatz \cite{Mezincescu90}
in the same manner as for the six-vertex model in Section 3.
Leaving out the
details, we find that the eigenvalue in the inhomogeneous case is given by
\begin{eqnarray}
{}_{\sigma}\Lambda(u,\boldomega) & = & \alpha(u) \frac{\sinh(2u+2\lambda)}
  {\sinh(2u+\lambda)}\prod_{i=1}^N \bar{a}(u+\omega_i)\bar{a}(u-\omega_i)
A(u) \nonumber\\
& + & \beta(u)\frac{\sinh(2u)}
 {\sinh(2u+\lambda)}\prod_{i=1}^N \bar{c}(u+\omega_i)\bar{c}(u-\omega_i) B(u),
\end{eqnarray}
where
\begin{eqnarray}
\alpha(u) & = & \sinh(u+\xi_--\smallfrac{\lambda}{2})
\sinh(u-\xi_+ + \smallfrac{\lambda}{2}),\\
\beta(u) & = & \sinh(u-\xi_-+\smallfrac{3\lambda}{2})
    \sinh(u+\xi_+ +\smallfrac{\lambda}{2}),\\
A(u) & = & \prod_{j=1}^M \frac{\sinh(u-u_j-\smallfrac{\lambda}{2})
             \sinh(u+u_j-\smallfrac{\lambda}{2})}
             {\sinh(u-u_j+\smallfrac{\lambda}{2})
             \sinh(u+u_j+\smallfrac{\lambda}{2})},\label{eqn:auzf}\\
B(u) & = & \prod_{j=1}^M \frac{\sinh(u-u_j+\smallfrac{3\lambda}{2})
             \sinh(u+u_j+\smallfrac{3\lambda}{2})}
             {\sinh(u-u_j+\smallfrac{\lambda}{2})
             \sinh(u+u_j+\smallfrac{\lambda}{2})}.
\end{eqnarray}

The eigenvalue for the original transfer matrix $t(u,\boldomega)$
is related to that for the auxiliary one by
\begin{equation}
\Lambda(u,\boldomega) = \gamma(u)\; {}_{\sigma}\Lambda(u-\smallfrac{\lambda}
  {2},\boldomega) \; {}_{\sigma}\Lambda(u+\smallfrac{\lambda}{2},\boldomega)
  + \delta(u)
\end{equation}
where we have defined
\begin{eqnarray}
\gamma(u) &=& \frac{\sinh(2u)\sinh(2u+2\lambda)}{\sinh(2u+\lambda)^2},
\nonumber\\
\delta(u) & = & - \alpha(u-\smallfrac{\lambda}{2})
   \beta(u+\smallfrac{\lambda}{2})\frac{\sinh(2u+3\lambda)\sinh(2u-\lambda)}
   {\sinh(2u+\lambda)^2}
 \prod_{i=1}^N  \bar{e}(u+\omega_i) \bar{e}(u-\omega_i),\nonumber\\
\bar{e}(u) & = & \sinh(u+2 \lambda)\sinh(u- \lambda).
\end{eqnarray}
The Bethe ansatz equations satisfied by the roots $u_j$ are the same ones
as for the auxiliary transfer matrix, namely for all $k$:
\begin{eqnarray}
\lefteqn{\frac{\sinh(u_k + \xi_- -\lambda)\sinh(u_k-\xi_+)}
  {\sinh(u_k - \xi_- +\lambda)\sinh(u_k+\xi_+)}
 \prod_{i=1}^N \frac{\sinh(u_k - \omega_i + \lambda)
   \sinh(u_k + \omega_i + \lambda)}{\sinh(u_k - \omega_i - \lambda)
   \sinh(u_k + \omega_i - \lambda)}}\hspace{120pt}\nonumber\\
& = & \prod_{j \neq k}^{M}\frac{\sinh(u_k-u_j+\lambda)\sinh(u_k+u_j+\lambda)}
    {\sinh(u_k-u_j-\lambda)\sinh(u_k+u_j-\lambda)}.
\end{eqnarray}
When $\omega_i=0$ these are the Bethe ansatz equations which determine the
energy spectrum for the spin-$1$ XXZ chain with open boundaries
\cite{Mezincescu90}. If, in addition, we choose $\xi_{\pm} \rightarrow
- \infty$ then the spin chain has $U_q(su(2))$-invariance \cite{Mezincescu90}.

With the alternating inhomogeneities $\boldomega=\boldsp(u)$, the transfer
matrix eigenvalue evaluates to
\begin{eqnarray}
\Lambda(u,\boldsp(u)) & = & \alpha(u-\smallfrac{\lambda}{2})\alpha(u+
   \smallfrac{\lambda}{2}) \sinh(2u+2\lambda) \sinh(2u+3\lambda)\nonumber\\
&  & \times \;   \bar{a}(2u)^{N-1} \rho(0)^{N/2} A(u).
\label{eqn:eigzf}
\end{eqnarray}
Once again, only one term survives. This seems to be a generic feature of
all vertex models on the lattice ${\cal L}$, as the coordinate Bethe ansatz
would also confirm. By substituting the expressions for the $R$- and
$K$-matrices into (\ref{eqn:lr}),
the left and right boundaries $l^b_a(u)$ and $r_a^b(u)$ are found to be
\begin{eqnarray}
l_1^1(u) & = & \sinh(2u+2\lambda)\sinh(2u+3\lambda)\sinh(\xi_+ -u)
   \sinh(\xi_+ - \lambda - u),\nonumber\\
l_2^2(u) & = & l_1^1(u) \frac{\sinh(\xi_+ - \lambda +u)}
    {\sinh(\xi_+ - \lambda -u)},\nonumber\\
l_3^3(u) & = & l_1^1(u) \frac{\sinh(\xi_+ + u)\sinh(\xi_+ - \lambda +u)}
    {\sinh(\xi_+ - u)\sinh(\xi_+ - \lambda - u)},\nonumber\\
r_1^1(u) & = & \sinh(u+\xi_-)\sinh(u-\lambda+\xi_-),\nonumber\\
r_2^2(u) & = & r_1^1(u) \frac{\sinh(\xi_- - u)}{\sinh(u+\xi_-)},\nonumber\\
r_3^3(u) & = & r_1^1(u) \frac{\sinh(u - \xi_-)\sinh(u+\lambda-\xi_-)}
  {\sinh(u+\xi_-)\sinh(u-\lambda+\xi_-)}.
\label{eqn:bozf}
\end{eqnarray}
It is then straightforward to show that $l_1^1(u) r_1^1 (u) =
\sinh(2u+2\lambda)\sinh(2u+3\lambda)\alpha(u-\smallfrac{\lambda}{2})
\alpha(u+\smallfrac{\lambda}{2})$. Thus the eigenvalue expression for the
diagonal-to-diagonal transfer matrix $t_D(u)$ is simply (cf.\ equations
(\ref{eqn:eigzf}) and (\ref{eqn:key}))
\begin{equation}
\Lambda_D(u) = A(u/2),
\end{equation}
where $A(u)$ is defined in (\ref{eqn:auzf}),
if we normalize all Boltzmann weights such that $w_{14}=w_1^L=w_1^R=1$.
The corresponding Bethe ansatz equations are given by
\begin{eqnarray}
\lefteqn{\frac{\sinh(u_k + \xi_- -\lambda)\sinh(u_k-\xi_+)}
  {\sinh(u_k - \xi_- +\lambda)\sinh(u_k+\xi_+)}
  \left[\frac{\sinh(u_k-\smallfrac{u}{2}+\lambda)
    \sinh(u_k+\smallfrac{u}{2}+\lambda)}
   {\sinh(u_k-\smallfrac{u}{2}-\lambda)\sinh(u_k+\smallfrac{u}{2}-\lambda)}
  \right]^N}\hspace{120pt}\nonumber\\
& = & \prod_{j \neq k}^{M}\frac{\sinh(u_k-u_j+\lambda)\sinh(u_k+u_j+\lambda)}
    {\sinh(u_k-u_j-\lambda)\sinh(u_k+u_j-\lambda)}.
\end{eqnarray}
The case where $\xi_-=u/2 + 2 \lambda$ and $\xi_+=-u/2-\lambda$ has previously
been obtained via the coordinate Bethe ansatz \cite{Batchelor93b}.

\section{The $A^{(1)}_2$ vertex model}
\setcounter{equation}{0}

The $R$-matrix associated with the affine  algebra $A^{(1)}_2$
in the vector representation is given by
\cite{Cherednik80,Babelon81}
\begin{equation}
R(u)=\left(\begin{array}{ccc|ccc|ccc}
    a & & & & & & & & \\
    & b & & c & & & & &\\
    & & b & & & & c & &\\ \hline
    & d & & b & & & & &\\
    & & & & a & & & &\\
    & & & & & b & & c &\\ \hline
    & & d & & & & b & &\\
    & & & & & d & & b &\\
    & & & & & & & & a \end{array} \right),
\end{equation}
where
\begin{eqnarray*}
  a(u) & = & \sinh(u+\lambda),\\
  b(u) & = & \sinh(u),\\
  c(u) & = & e^u\sinh(\lambda),\\
  d(u) & = & e^{-u}\sinh(\lambda).
\end{eqnarray*}
This $R$-matrix has the properties of regularity and unitarity, with
$\rho(u)=\sinh(\lambda+u)\sinh(\lambda-u)$, $PT$-symmetry and a weak version
of crossing-symmetry \cite{Mezincescu92a,deVega93}
\begin{equation}
    \left(\left(\left(R_{12}(u)^{t_2}\right)^{-1}\right)^{t_2}\right)^{-1}=
   f(u) \stackrel{2}{M} R_{12}(u+2\eta) \stackrel{2}{M^{-1}},
\label{eqn:weak}
\end{equation}
with $\eta=3\lambda/2$, $M=\diag(e^{2\lambda},1,e^{-2\lambda})$ and scalar
factor
\begin{equation}
f(u)=\frac{\sinh(u)\sinh(u+3\lambda)}{\sinh(u+2\lambda)\sinh(u+4\lambda)}.
\end{equation}
With the property (\ref{eqn:weak}) replacing crossing-symmetry, the
condition (\ref{eqn:k2}) on $K^+(u)$ for
integrability remains valid \cite{Mezincescu92a},
with the automorphism (\ref{eqn:auto}) again holding.
It is easy to see that $K^-(u) = 1$ is a
solution of (\ref{eqn:k1}) for the present model, as with all the models
considered here. The most general diagonal solution
\cite{deVega93} for $K^-(u)$ is given by $K^-(u)=K(u,\xi_-,\l_-)$ where
\begin{equation}
K_{aa}(u,\xi,l)=\left\{\begin{array}{lll}
      \sinh(\xi - u) e^u & &1\leq a \leq l\\
      \sinh(\xi+u) e^{-u} & &l+1 \leq a \leq 3
  \end{array}\right.,
\end{equation}
where  $\xi$ is arbitrary and $l$ can be either $1,2$ or $3$.
By (\ref{eqn:auto}) solutions for $K^+(u)$ follow, with corresponding
$\xi_+$ and $l_+$.

The most general commuting transfer matrix $t(u,\boldomega)$ then depends
on the four parameters $\xi_{\pm}, l_{\pm}$. This transfer matrix has recently
been diagonalized with
 the nested Bethe ansatz \cite{deVega94b}, generalizing an
earlier result \cite{deVega94a} valid for the case $K^-(u)=1$ and $K^+(u)=M$,
which leads to an $U_q(su(3))$-invariant spin chain \cite{Mezincescu91c}.
In fact these authors
studied the whole family of $A_n^{(1)}, n\geq 1$ vertex models.
We will only be interested here in the three cases where
 $l_-=l_+$. Note that the case $l_+=l_-=3$ is actually equivalent to the
$U_q(su(3))$-invariant case, by rescaling the $K$-matrices.
The expression for the eigenvalue $\Lambda(u,
\boldomega)$ can be written as
\begin{eqnarray}
\Lambda(u,\boldomega) & = & \alpha(u)
  \frac{\sinh(2u+3\lambda)}{\sinh(2u+\lambda)}
  \prod_{i=1}^{N}a(u+\omega_i)
  a(u-\omega_i) A(u) \nonumber\\
  & & + \;\beta(u) \frac{\sinh(2u)\sinh(2u+3\lambda)}
   {\sinh(2u+\lambda)\sinh(2u+2\lambda)} \prod_{j=1}^{N}
  b(u+\omega_j)b(u-\omega_j) B(u) C(u) \nonumber\\
  & &  + \;\gamma(u) \frac{\sinh(2u)}{\sinh(2u+2\lambda)}
  \prod_{i=1}^{N} b(u+\omega_i)b(u-\omega_i) D(u)
\label{eqn:eg}
\end{eqnarray}
where
\begin{eqnarray}
A(u) & = & \prod_{j=1}^{M_1} \frac{\sinh(u+ \mu_j^{(1)} -
     \smallfrac{\lambda}{2})
              \sinh(u- \mu_j^{(1)} - \smallfrac{\lambda}{2})}
             {\sinh(u+ \mu_j^{(1)} + \smallfrac{\lambda}{2})
              \sinh(u- \mu_j^{(1)} + \smallfrac{\lambda}{2})},
\label{eqn:aua21}\\
B(u) & = & \prod_{j=1}^{M_1} \frac{\sinh(u+ \mu_j^{(1)} +
      \smallfrac{3\lambda}{2})
              \sinh(u- \mu_j^{(1)} + \smallfrac{3 \lambda}{2})}
             {\sinh(u+ \mu_j^{(1)} + \smallfrac{\lambda}{2})
              \sinh(u- \mu_j^{(1)} + \smallfrac{\lambda}{2})},\\
C(u) & = & \prod_{j=1}^{M_2} \frac{\sinh(u+ \mu_j^{(2)} +
      \smallfrac{\lambda}{2})
              \sinh(u- \mu_j^{(2)} - \smallfrac{\lambda}{2})}
             {\sinh(u+ \mu_j^{(2)} + \smallfrac{3\lambda}{2})
              \sinh(u- \mu_j^{(2)} + \smallfrac{\lambda}{2})},\\
D(u) & = & \prod_{j=1}^{M_2} \frac{\sinh(u+ \mu_j^{(2)} +
       \smallfrac{5\lambda}{2})
              \sinh(u- \mu_j^{(2)} + \smallfrac{3\lambda}{2})}
             {\sinh(u+ \mu_j^{(2)} + \smallfrac{3\lambda}{2})
              \sinh(u- \mu_j^{(2)} + \smallfrac{\lambda}{2})}.
\end{eqnarray}
The (nested)
Bethe ansatz equations follow from analyticity of $\Lambda(u,\boldomega)$
and are given by
\begin{eqnarray}
\lefteqn{\frac{\alpha(\mu_k^{(1)}-\smallfrac{\lambda}{2})}
              {\beta(\mu_k^{(1)}-\smallfrac{\lambda}{2})}
   \prod_{i=1}^N \frac{\sinh(\mu_k^{(1)}+\omega_i+\smallfrac{\lambda}{2})
                    \sinh(\mu_k^{(1)}-\omega_i+\smallfrac{\lambda}{2})}
                   {\sinh(\mu_k^{(1)}+\omega_i-\smallfrac{\lambda}{2})
                    \sinh(\mu_k^{(1)}-\omega_i-\smallfrac{\lambda}{2})}}
\hspace{60pt} \nonumber\\
&= & \prod_{j\neq k}^{M_1} \frac{\sinh(\mu_k^{(1)} + \mu_j^{(1)} + \lambda)
                    \sinh(\mu_k^{(1)} - \mu_j^{(1)} + \lambda)}
                   {\sinh(\mu_k^{(1)} + \mu_j^{(1)} - \lambda)
                    \sinh(\mu_k^{(1)} - \mu_j^{(1)} - \lambda)}\nonumber\\
 & & \times \; \prod_{j=1}^{M_2}\frac{\sinh(\mu_k^{(1)} + \mu_j^{(2)}-
                                 \smallfrac{\lambda}{2})
                    \sinh(\mu_k^{(1)} - \mu_j^{(2)}-\smallfrac{\lambda}{2})}
                   { \sinh(\mu_k^{(1)} + \mu_j^{(2)}+\smallfrac{\lambda}{2})
                     \sinh(\mu_k^{(1)} - \mu_j^{(2)}+\smallfrac{\lambda}{2})},
\\
\frac{\beta(\mu_k^{(2)}-\smallfrac{\lambda}{2})}
               {\gamma(\mu_k^{(2)}-\smallfrac{\lambda}{2})}
 & = & \prod_{j\neq k}^{M_2}  \frac{\sinh(\mu_k^{(2)} + \mu_j^{(2)} + \lambda)
                    \sinh(\mu_k^{(2)} - \mu_j^{(2)} + \lambda)}
                   {\sinh(\mu_k^{(2)} + \mu_j^{(2)} - \lambda)
                    \sinh(\mu_k^{(2)} - \mu_j^{(2)} - \lambda)}\nonumber\\
 & &  \times \;  \prod_{j=1}^{M_1}\frac{\sinh(\mu_k^{(2)} + \mu_j^{(1)}-
                                \smallfrac{\lambda}{2})
                    \sinh(\mu_k^{(2)} - \mu_j^{(1)}- \smallfrac{\lambda}{2})}
                    {\sinh(\mu_k^{(2)} + \mu_j^{(1)}+ \smallfrac{\lambda}{2})
                    \sinh(\mu_k^{(2)} - \mu_j^{(1)}+ \smallfrac{\lambda}{2})},
\label{eqn:ba2}
\end{eqnarray}
for all $k$.
For the three types of boundaries under consideration
the scalar factors $\alpha$, $\beta$ and $\gamma$ are  given by

\noindent (i) $l_-=l_+=1$:
\begin{eqnarray}
\alpha(u) & = & e^{\smallfrac{\lambda}{2}}\sinh(\xi_- - u)\sinh(\xi_+ + u-
     \smallfrac{\lambda}{2}) \nonumber\\
\beta(u)=\gamma(u) & = & e^{\smallfrac{\lambda}{2}}\sinh(\xi_- + u + \lambda)
     \sinh(\xi_+ - u- \smallfrac{3\lambda}{2}),
\end{eqnarray}

\noindent (ii) $l_-=l_+=2$:
\begin{eqnarray}
\alpha(u) =\beta(u)
   & = & e^{-\smallfrac{\lambda}{2}}\sinh(\xi_- - u)\sinh(\xi_+ + u+
     \smallfrac{\lambda}{2}) \nonumber\\
\gamma(u) & = & e^{-\smallfrac{\lambda}{2}}\sinh(\xi_- + u + 2\lambda)
     \sinh(\xi_+ - u- \smallfrac{3\lambda}{2})
\end{eqnarray}

\noindent (iii) $l_-=l_+=3$:
\begin{equation}
\alpha(u) = \beta(u) = \gamma(u)
     =  e^{-\smallfrac{3\lambda}{2}}\sinh(\xi_- - u)\sinh(\xi_+ + u+
     \smallfrac{3\lambda}{2}).
\end{equation}
These expressions for $\alpha(u)$, $\beta(u)$ and $\gamma(u)$ follow from
the results in Ref. \cite{deVega94b}, after a shift in $\xi_+$ necessary to
match
up with the different parametrization of $K^+(u)$ used therein.

We now specialize the inhomogeneities $\boldomega$ to $\boldsp(u)$. As usual,
only the first term in the eigenvalue expression (\ref{eqn:eg}) survives:
\begin{equation}
\Lambda(u,\boldsp(u)) = \alpha(u) \sinh(2u+3\lambda) a(2u)^{N-1}
  \rho(0)^{N/2} A(u).
\label{eqn:eiga21}
\end{equation}
The right and left boundary weights $r_a^b(u)$ and $l_a^b(u)$
follow from the definitions (\ref{eqn:lr}),
and for the three cases of interest read:

\noindent (i) $l_-=l_+=1$:
\begin{eqnarray}
l_1^1(u) & = & \sinh(2u+3\lambda)\sinh(u+\xi_+-\smallfrac{\lambda}{2})
   e^{\lambda/2 -u},\nonumber\\
l_2^2(u) = l_3^3(u) & = & l^1_1(u) e^{2u},\nonumber\\
r_1^1(u) & = & \sinh(\xi_- - u) e^u,\nonumber\\
r_2^2(u) = r_3^3(u)
  & = & r_1^1(u) \frac{\sinh(\xi_- +u)}{\sinh(\xi_- -u)} e^{-2u},
\label{eqn:boa211}
\end{eqnarray}

\noindent (ii) $l_-=l_+=2$:
\begin{eqnarray}
l_1^1(u) = l_2^2(u)
   & = & \sinh(2u+3\lambda)\sinh(u+\xi_++\smallfrac{\lambda}{2})
   e^{-\lambda/2 -u},\nonumber\\
l_3^3(u) & = & e^{2u}\frac{\sinh(\smallfrac{\lambda}{2}+\xi_+ - u)}
    {\sinh(\smallfrac{\lambda}{2} + \xi_+ + u)},\nonumber\\
r_1^1(u)=r_2^2(u)  & = & \sinh(\xi_- - u) e^u,\nonumber\\
r_3^3(u) & = & r_1^1(u) \frac{\sinh(\xi_- +u)}{\sinh(\xi_- -u)} e^{-2u},
\label{eqn:boa212}
\end{eqnarray}

\noindent (iii) $l_-=l_+=3$:
\begin{eqnarray}
l_1^1(u) = l_2^2(u) = l_3^3(u)
   & = & \sinh(2u+3\lambda)\sinh(u+\xi_++\smallfrac{3\lambda}{2})
   e^{-3\lambda/2 -u},\nonumber\\
r_1^1(u) = r_2^2(u) = r_3^3(u) & = & \sinh(\xi_- - u) e^u.
\label{eqn:boa213}
\end{eqnarray}
{}From these we find that $l_1^1(u) r_1^1(u) = \alpha(u) \sinh(2u+3\lambda)$
for all cases. Therefore
if we normalize all the vertex weights such that $w_{14}=1=w^R_1=w^L_1$,
the eigenvalue for the diagonal-to-diagonal transfer matrix $t_D(u)$
becomes (cf.\ (\ref{eqn:key}) and (\ref{eqn:eiga21}))
simply $\Lambda_D(u) = A\left(\smallfrac{u}{2}\right)$, with $A(u)$ defined
in (\ref{eqn:aua21}).
The corresponding Bethe ansatz equations are given by (\ref{eqn:ba2}) and
\begin{eqnarray}
\lefteqn{\frac{\alpha(\mu_k^{(1)}-\smallfrac{\lambda}{2})}
              {\beta(\mu_k^{(1)}-\smallfrac{\lambda}{2})}
         \left[\frac{\sinh(\mu_k^{(1)}+\smallfrac{u}{2}+\smallfrac{\lambda}{2})
                    \sinh(\mu_k^{(1)}-\smallfrac{u}{2}+\smallfrac{\lambda}{2})}
                   {\sinh(\mu_k^{(1)}+\smallfrac{u}{2}-\smallfrac{\lambda}{2})
                    \sinh(\mu_k^{(1)}-\smallfrac{u}{2}
   -\smallfrac{\lambda}{2})}\right]^N}
 \hspace{20pt} \nonumber\\
& =&\prod_{j\neq k}^{M_1} \frac{\sinh(\mu_k^{(1)} + \mu_j^{(1)} + \lambda)
                    \sinh(\mu_k^{(1)} - \mu_j^{(1)} + \lambda)}
                   {\sinh(\mu_k^{(1)} + \mu_j^{(1)} - \lambda)
                    \sinh(\mu_k^{(1)} - \mu_j^{(1)} - \lambda)}\nonumber\\
& &  \times \; \prod_{j=1}^{M_2}\frac{\sinh(\mu_k^{(1)} + \mu_j^{(2)}-
                                 \smallfrac{\lambda}{2})
                    \sinh(\mu_k^{(1)} - \mu_j^{(2)}-\smallfrac{\lambda}{2})}
                   { \sinh(\mu_k^{(1)} + \mu_j^{(2)}+\smallfrac{\lambda}{2})
                     \sinh(\mu_k^{(1)} - \mu_j^{(2)}+\smallfrac{\lambda}{2})}.
\end{eqnarray}

\section{The $A^{(2)}_2$ vertex model}
\setcounter{equation}{0}
The $R$-matrix for the Izergin-Korepin  \cite{Izergin81} or
$A^{(2)}_2$  model \cite{Jimbo86,Bazhanov87} is given by
\begin{equation}
R(u)=\left(\begin{array}{ccc|ccc|ccc}
    c & & & & & & & & \\
    & b & & e & & & & &\\
    & & d & & g &  & f & &\\ \hline
    & \bar{e} & & b & & & & &\\
    & & \bar{g}& & a & & g & &\\
    & & & & & b & & e &\\ \hline
    & & & & & & & &\\[-10pt]
    & & \bar{f} & & \bar{g} & & d & &\\
    & & & & & \bar{e} & & b &\\
    & & & & & & & & c \end{array} \right),
\end{equation}
where
\begin{eqnarray*}
a(u) & = & \sinh(u-3\lambda) - \sinh(5\lambda) + \sinh(3\lambda) +
   \sinh(\lambda),\\
b(u) & = & \sinh(u-3\lambda) + \sinh(3\lambda),\\
c(u) & = & \sinh(u-5\lambda) + \sinh(\lambda),\\
d(u) & = & \sinh(u-\lambda) + \sinh(\lambda),\\
e(u) & = & -2e^{-\smallfrac{u}{2}}\sinh(2\lambda)\cosh(\smallfrac{u}{2}-
   3\lambda),\\
\bar{e}(u) & = & -2 e^{\smallfrac{u}{2}}\sinh(2\lambda)\cosh(\smallfrac{u}{2}-
   3\lambda),\\
f(u) & = & -2 e^{-u+2\lambda}\sinh(\lambda)\sinh(2\lambda) - e^{-\lambda}
   \sinh(4\lambda),\\
\bar{f}(u) & = & 2e^{u-2\lambda}\sinh(\lambda)\sinh(2\lambda) -
   e^{\lambda}\sinh(4\lambda),\\
g(u) & = & 2 e^{-\smallfrac{u}{2}+2\lambda}\sinh(\smallfrac{u}{2})
    \sinh(2\lambda),\\
\bar{g}(u) & = & -2 e^{\smallfrac{u}{2}-2\lambda}\sinh(\smallfrac{u}{2})
    \sinh(2\lambda).\\
\end{eqnarray*}
This $R$-matrix has the properties of regularity and unitarity, with
$\rho(u)=[\sinh(\lambda)+\sinh(u-5\lambda)]
[\sinh(\lambda)-\sinh(u+5\lambda)]$.
It is also $PT$-symmetric and crossing-symmetric, with
$\eta=-6\lambda-\I\pi$ and
\begin{equation}
   M=\diag(e^{2\lambda},1,e^{-2\lambda}).
\end{equation}

Diagonal solutions for $K^-(u)$ have been obtained in \cite{Mezincescu91b}.
It turns out that there are only three solutions, being $K^-(u)=1$,
$K^-(u) = \Gamma^+(u)$ and $K^-(u) = \Gamma^-(u)$, with
\begin{equation}
\Gamma^{\pm}(u)=\left(\begin{array}{ccc}
 e^{-u}\psi^{\pm}(u) & &\\
 &      \phi^{\pm}(u) & \\
 & & e^{u} \psi^{\pm}(u)
\end{array} \right),
\label{eqn:kminus1}
\end{equation}
where we have defined
\begin{eqnarray}
\psi^{\pm}(u) & = & \cosh(\smallfrac{u}{2}-3\lambda) \pm  \I
\sinh(\smallfrac{u}{2})\\
\phi^{\pm}(u) & = &
\cosh(\smallfrac{u}{2}+3\lambda)\mp \I \sinh(\smallfrac{u}{2}).
\end{eqnarray}
There is an absence of free parameters in these solutions, in contrast to the
previous models. By the automorphism (\ref{eqn:auto}),
three solutions for $K^+(u)$ follow.
We have thus nine possibilities for the commuting transfer matrix
$t(u,\boldomega)$. We will only be interested in the three cases
\begin{enumerate}
\item[(i)] $K^-(u)=1$, $K^+(u)=M$,
\item[(ii)] $K^-(u)=\Gamma^+(u)$,
$K^+(u)=\Gamma^+(-u+6\lambda+\I \pi)^t M$,
\item[(iii)] $K^-(u)=\Gamma^-(u)$,
$K^+(u)=\Gamma^-(-u+6\lambda+\I \pi)^t M$.
\end{enumerate}

The transfer matrix for case (i), whose corresponding spin-chain is
$U_q(su(2))$-invariant \cite{Mezincescu91c},
has been diagonalized by the analytic
Bethe ansatz in Ref. \cite{Mezincescu92b} for the homogeneous
case $\omega_i=0$. It is not
difficult to introduce the inhomogeneities, and the result for the eigenvalue
can be written as
\begin{eqnarray}
\Lambda(u,\boldomega) & = & \prod_{i=1}^N c(u+\omega_i)c(u-\omega_i)
  \frac{\sinh(u-6\lambda)\cosh(u-\lambda)}{\sinh(u-2\lambda)\cosh(u-3\lambda)}
  A(u) \nonumber\\
& + & \prod_{i=1}^N b(u+\omega_i)b(u-\omega_i)
 \frac{\sinh(u)\sinh(u-6\lambda)}{\sinh(u-2\lambda)\sinh(u-4\lambda)}
  B(u) \nonumber\\
& + & \prod_{i=1}^N d(u+\omega_i)d(u-\omega_i)
 \frac{\sinh(u)\cosh(u-5\lambda)}{\sinh(u-4\lambda)\cosh(u-3\lambda)}
 C(u),
\label{eqn:mn}
\end{eqnarray}
with
\begin{eqnarray}
A(u) & = & \prod_{j=1}^{m} \frac{\sinh(\smallfrac{1}{2}(u+u_j) +\lambda)
   \sinh(\smallfrac{1}{2}(u-u_j) +\lambda)}
   {\sinh(\smallfrac{1}{2}(u+u_j) -\lambda)
        \sinh(\smallfrac{1}{2}(u-u_j) -\lambda)}\label{eqn:aua22}\\
B(u) & = & \prod_{j=1}^{m} \frac{\sinh(\smallfrac{1}{2}(u+u_j) -3\lambda)
     \sinh(\smallfrac{1}{2}(u-u_j) -3\lambda)}
      {\sinh(\smallfrac{1}{2}(u+u_j) -\lambda)
       \sinh(\smallfrac{1}{2}(u-u_j) -\lambda)}\nonumber\label{eqn:b}\\
     &  & \times \; \prod_{j=1}^{m}\frac{\cosh(\smallfrac{1}{2}(u+u_j))
           \cosh(\smallfrac{1}{2}(u-u_j))}
       {\cosh(\smallfrac{1}{2}(u+u_j)-2\lambda)
           \cosh(\smallfrac{1}{2}(u-u_j)-2\lambda)}\\
C(u) & = & \prod_{j=1}^{m}\frac{\cosh(\smallfrac{1}{2}(u+u_j) -4\lambda)
   \cosh(\smallfrac{1}{2}(u-u_j) -4\lambda)}
   {\cosh(\smallfrac{1}{2}(u+u_j) -2\lambda)
         \cosh(\smallfrac{1}{2}(u-u_j) -2\lambda)}.
\label{eqn:c}
\end{eqnarray}

For the above case, quantum group invariance was used to determine properties
of the transfer matrix eigenvalue essential for the analytic Bethe ansatz to
be pushed through. For the non-quantum group invariant cases, the analogous
calculations have not been carried out.
Nevertheless, we expect the eigenvalue expression to be simply modified to
\begin{eqnarray}
\Lambda(u,\boldomega) & = & \prod_{i=1}^N c(u+\omega_i)c(u-\omega_i)
 \alpha(u) A(u)
 +  \prod_{i=1}^N b(u+\omega_i)b(u-\omega_i)
  \beta(u) B(u) \nonumber\\
& + & \prod_{i=1}^N d(u+\omega_i)d(u-\omega_i)
 \gamma(u) C(u),
\label{eqn:eiga22}
\end{eqnarray}
with $A(u)$, $B(u)$ and $C(u)$ defined as in (\ref{eqn:aua22}),
(\ref{eqn:b}) and (\ref{eqn:c}) and the
boundary-dependent terms $\alpha(u)$,
$\beta(u)$ and $\gamma(u)$ are to be
determined by the action of $t(u,\boldomega)$
on the pseudovacuum $|\Omega\rangle$ (state with all arrows up).
By a straightforward calculation -- e.g.\ by
graphical means (Fig. 6) --  we find them to be given by
\begin{eqnarray}
\alpha & = & K_{11}^-\left\{ K_{11}^+ + \frac{K_{22}^+ \bar{e}^2}
  {c^2-b^2} + \frac{K_{33}^+}{c^2-d^2}\left( \bar{f}^2 + \frac{\bar{g}^2
  \bar{e}^2}{c^2-b^2}\right)\right\}\nonumber\\
\beta & = & K_{22}^+ K_{22}^- - \frac{K_{22}^+K_{11}^-\bar{e}^2}{c^2-b^2}-
   \frac{K_{33}^+K_{22}^-\bar{g}^2}{d^2-b^2} - \frac{K_{33}^+K_{11}^-
 \bar{g}^2\bar{e}^2}{(c^2-b^2)(b^2-d^2)}\nonumber\\
\gamma & = & K_{33}^+\left\{ K_{33}^- + \frac{K_{22}^-\bar{g}^2}{d^2-b^2} -
  \frac{K_{11}^-}{c^2-d^2}\left( \bar{f}^2 - \frac{\bar{g}^2\bar{e}^2}
  {b^2-d^2}\right) \right\}.
\label{eqn:mnev}
\end{eqnarray}
Naturally, for case (i) the result (\ref{eqn:mn}) is recovered. For the
other cases (\ref{eqn:mnev}) yields

\noindent (ii):
\begin{eqnarray}
\alpha(u) & = & \frac{(\psi^+)^2}{c(2u)}\xi^+(u)\sinh(u-6\lambda),\nonumber\\
\beta(u) &=& \alpha(u) \frac{\sinh(u)}{\sinh(u-4\lambda)}
\frac{\cosh(\lambda)+\I\sinh(2\lambda-u)}{\cosh(\lambda)-\I\sinh(2\lambda-u)},
\nonumber\\
\gamma(u) & = & \alpha(u)\frac{\xi^-(u)}{\xi^+(u)}
   \frac{\sinh(u)\sinh(u-2\lambda)}{\sinh(u-6\lambda)\sinh(u-4\lambda)},
\end{eqnarray}
\noindent (iii):
\begin{eqnarray}
\alpha(u) & = & -\frac{(\psi^-)^2}{c(2u)}\xi^-(u)\sinh(u-6\lambda),\nonumber\\
\beta(u) &=& \alpha(u) \frac{\sinh(u)}{\sinh(u-4\lambda)}
\frac{\cosh(\lambda)-\I\sinh(2\lambda-u)}{\cosh(\lambda)+\I\sinh(2\lambda-u)},
\nonumber\\
\gamma(u) & = & \alpha(u)\frac{\xi^+(u)}{\xi^-(u)}
   \frac{\sinh(u)\sinh(u-2\lambda)}{\sinh(u-6\lambda)\sinh(u-4\lambda)},
\end{eqnarray}
where we have defined
\begin{equation}
\xi^{\pm}(u) = 2 \left(\cosh(3\lambda-u) \pm i \sinh(2\lambda)\right).
\end{equation}
The Bethe ansatz equations associated with the eigenvalue expression
(\ref{eqn:eiga22}) are  given by
\begin{eqnarray}
\lefteqn{
\prod_{i=1}^N \left[\frac{c(u_k+2\lambda+\omega_i)c(u_k+2\lambda-\omega_i)}
 {b(u_k+2\lambda+\omega_i)b(u_k+2\lambda-\omega_i)}\right]
 \frac{\alpha(u_k+2\lambda)}
 {\beta(u_k+2\lambda)}\frac{\sinh(u_k+2\lambda)\cosh(u_k-\lambda)}
  {\sinh(u_k-2\lambda)\cosh(u_k+\lambda)}}\hspace{90pt}\nonumber\\
&= & \prod_{j\neq k}^{m} \frac{\sinh[\smallfrac{1}{2}(u_k+u_j)-2\lambda]
 \sinh[\smallfrac{1}{2}(u_k-u_j)-2\lambda]}
{\sinh[\smallfrac{1}{2}(u_k+u_j)+2\lambda]
 \sinh[\smallfrac{1}{2}(u_k-u_j)+2\lambda]} \nonumber\\
& & \times \; \frac{
\cosh[\smallfrac{1}{2}(u_k+u_j)+\lambda]
 \cosh[\smallfrac{1}{2}(u_k-u_j)+\lambda]}
{\cosh[\smallfrac{1}{2}(u_k+u_j)-\lambda]
 \cosh[\smallfrac{1}{2}(u_k-u_j)-\lambda]}.
\label{eqn:bet}
\end{eqnarray}

Following the general procedure laid down in Section 2,
we set $\boldomega=\boldsp(u)$ in (\ref{eqn:eiga22}). As usual, only the first
term survives:
\begin{equation}
\Lambda(u,\boldsp(u))  =  c(2u)^N c(0)^N \alpha(u) A(u).
\end{equation}
{}From the definitions (\ref{eqn:lr}) the left and right boundary weights
$l_a^b(u)$ and $r_a^b(u)$ are found to be given by

\noindent (i):
\begin{eqnarray}
l_1^1(u)  =  l_2^2(u)  =  l_3^3(u) & = & 2 \sinh(u-6\lambda)\cosh(u-\lambda)
\nonumber\\
r_1^1(u)  =  r_2^2(u) = r_3^3(u) & = & 1,
\label{eqn:boa221}
\end{eqnarray}
\noindent (ii):
\begin{eqnarray}
l_1^1(u) & = & \psi^+(u) \sinh(u-6\lambda)e^u \xi^+(u)\nonumber\\
l_2^2(u) & = & l_1^1(u) \frac{\phi^+(u)}{\psi^+(u)} e^{-u}\nonumber\\
l_3^3(u) & = & l_1^1(u)  e^{-2u}\nonumber\\
r_1^1(u) & = & \psi^+(u)  e^{-u}\nonumber\\
r_2^2(u) & = & r_1^1(u) \frac{\phi^+(u)}{\psi^+(u)} e^{u}\nonumber\\
r_3^3(u) & = & r_1^1(u) e^{2u},
\label{eqn:boa222}
\end{eqnarray}
\noindent (iii):
\begin{eqnarray}
l_1^1(u) & = & -\psi^-(u) \sinh(u-6\lambda)e^u \xi^-(u)\nonumber\\
l_2^2(u) & = & l_1^1(u) \frac{\phi^-(u)}{\psi^-(u)} e^{-u}\nonumber\\
l_3^3(u) & = & l_1^1(u)  e^{-2u}\nonumber\\
r_1^1(u) & = & \psi^-(u)  e^{-u}\nonumber\\
r_2^2(u) & = & r_1^1(u) \frac{\phi^-(u)}{\psi^-(u)} e^{u}\nonumber\\
r_3^3(u) & = & r_1^1(u) e^{2u}.
\label{eqn:boa223}
\end{eqnarray}
For all three cases, we have
$l_1^1(u) r_1^1(u) = c(2u) \alpha(u)$.
Hence, by normalizing all bulk weights
such that $w_{14}=1$ and boundary weights such that $w^R_1=w^L_1=1$, the
eigenvalue for the diagonal-to-diagonal transfer matrix $t_D(u)$ is simply
\begin{equation}
\Lambda_D(u)  =  A\left(\smallfrac{u}{2}\right),
\label{eqn:eva22}
\end{equation}
where $A(u)$ is defined in (\ref{eqn:aua22}).
The corresponding Bethe ansatz equations are given by (\ref{eqn:bet}) with
the left hand side replaced by
\begin{equation}
\left[\frac{c(u_k+2\lambda+\smallfrac{u}{2})c(u_k+2\lambda-\smallfrac{u}{2})}
 {b(u_k+2\lambda+\smallfrac{u}{2})b(u_k+2\lambda-\smallfrac{u}{2})}\right]^N
  \left\{
 \frac{\alpha(u_k+2\lambda)}
 {\beta(u_k+2\lambda)}\frac{\sinh(u_k+2\lambda)\cosh(u_k-\lambda)}
  {\sinh(u_k-2\lambda)\cosh(u_k+\lambda)}\right\}.
\label{eqn:lh}
\end{equation}
For case (i), the term in braces
in (\ref{eqn:lh}) becomes unity (cf.\ (\ref{eqn:mn})), whereas
for cases (ii) and (iii) it simplifies to
\begin{equation}
  \left(
  \frac{\cosh[\smallfrac{1}{2}(u_k-\lambda + \smallfrac{\I \pi}{2})]}
       {\cosh[\smallfrac{1}{2}(u_k+\lambda - \smallfrac{\I \pi}{2})]}
  \right)^2
\end{equation}
and
\begin{equation}
  \left(
  \frac{\sinh[\smallfrac{1}{2}(u_k-\lambda + \smallfrac{\I \pi}{2})]}
       {\sinh[\smallfrac{1}{2}(u_k+\lambda - \smallfrac{\I \pi}{2})]}
  \right)^2
\label{eqn:bs}
\end{equation}
respectively.

\section{Loop models with integrable boundaries}
\setcounter{equation}{0}

Let ${\cal L}$ be the lattice depicted in Fig. 1. A loop model on ${\cal L}$
has partition function
\begin{equation}
Z_{\rm loop} = \sum_{{\cal G}} \rho_1^{m_1} \cdots \rho_{13}^{m_{13}} \; n^{P}
\end{equation}
where the sum is over all configurations ${\cal G}$ of non-intersecting closed
loops which cover some (or none) of the edges of ${\cal L}$. The possible
configurations at each vertex are shown in Fig. 10, with the one of type
$i$ carrying a Boltzmann weight $\rho_i$. In the configuration ${\cal G}$,
$m_i$ is the number of occurences of the vertex of type $i$ while $P$ is the
total number of closed loops of fugacity $n$.
Much work has been done on models with
periodic boundaries (see e.g.\ Refs. \cite{Nienhuis90,Warnaar93}),
where the boundary weights $\rho_{10}$ to $\rho_{13}$
are absent. In particular an $O(n)$ vector model can be mapped onto such a
loop model, with $n$ being precisely the loop fugacity in $Z_{\rm loop}$. The
$n \rightarrow 0$ limit of such an $O(n)$ model is especially
interesting because of its polymer interpretation \cite{deGennes79}.

%%%%%%%%%%%%%%%%%%  Figure 10 here  %%%%%%%%%%%%%%%%%%%%%
\begin{figure}[htb]
\epsfxsize = 12cm
\vbox{\vskip .8cm\hbox{\centerline{\epsffile{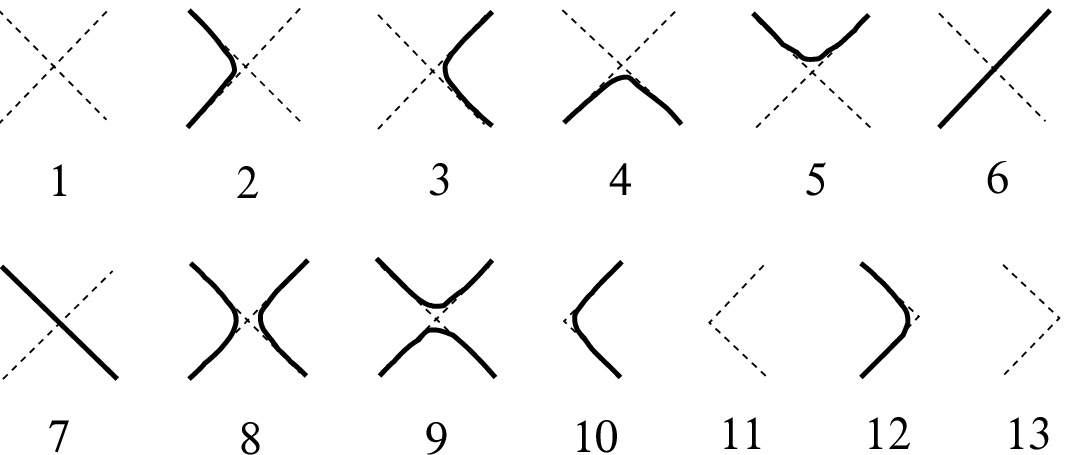}}}
%50%
\vskip .5cm \smallskip}
\caption{Allowed vertices for the loop model with partition function
$Z_{\rm loop}$. Vertex $i$ carries Boltzmann weight $\rho_i$.}
\end{figure}
%%%%%%%%%%%%%%%%%%% End figure 10   %%%%%%%%%%%%%%%%%%%%%%%

Instances where the loop model admits exact solutions are known. Such
important cases can be related to more well-known integrable vertex models
via the loop model-to-vertex model mapping \cite{Nienhuis90,Warnaar93}.
This involves first
assigning orientations to the loops in ${\cal G}$. The explicit closed
loop counting in $Z_{\rm loop}$, which is a non-local procedure, can be
eliminated by introducing a weight $s$ ($s^{-1}$) for each left (right) turn
along an oriented loop if $s$ is chosen so that $n=s^4+s^{-4}$. The loop
model partition function then becomes equivalent to a (three-state)
vertex model partition function (\ref{eqn:vp}),
whose weights are functions of $\rho_i$, $s$ and, in general,
gauge factors $a, b, c$ and $d$ (whose presence leave the partition
function unchanged). The relations between the vertex weights $w_i$,
$w_i^L$ and $w_i^R$ and the loop weights $\rho_i$  are
given explicitly by (refer to Figs. 3, 4  and 10)
\begin{equation}
\begin{array}{lllllll}
w_1 & = & \rho_1 & \hspace{20pt}
      &w_1^L & = & \rho_{10} s^{-1} \frac{1}{(ac)}\nonumber\\
w_2 & = & \rho_2 s (ac) & \hspace{20pt} &
  w_2^L & = & \rho_{11} \nonumber\\
w_3 & = & \rho_2 s^{-1} (bd) & \hspace{20pt}&
  w_3^L & = & \rho_{10} s^ \frac{1}{(bd)}\nonumber\\
\vdots & & \vdots  & \hspace{20pt}
  & w_1^R & = & \rho_{12} s (ac)\nonumber\\[5pt]
w_{18} &=& \left(\rho_8 s^{2} + \rho_9 s^{-2} \right) \frac{ac}{bd}
   & \hspace{20pt} & w_2^R & = & \rho_{13} \nonumber\\[5pt]
w_{19} &=& \left(\rho_8 s^{-2} + \rho_9 s^{2} \right) \frac{bd}{ac}&
   \hspace{20pt}& w_3^R & = & \rho_{12} s^{-1} (bd).
\end{array}
\label{eqn:necc}
\end{equation}
The question we want to pose is the following:
Given an integrable vertex model on ${\cal L}$, can a loop model
be found? For this to be possible, the vertex weights must satisfy
a set of constraints implied by the equations (\ref{eqn:necc}). The
constraints on the bulk weights $w_i$ are precisely those
for the model with periodic boundaries \cite{Nienhuis90,Warnaar93}.
For open boundaries, we see that the constraints
\begin{equation}
w^L_3 : w^L_1 \;\;=\;\; w^R_1 : w^R_3
\;\; = \;\;w_2 : w_3
\label{eqn:loop}
\end{equation}
on the boundary weights must also be satisfied. The important thing to note
about the criterion (\ref{eqn:loop}) is that it is gauge-independent.

\subsection{The $O(n)$ loop model}
To see how this works in practice, we look at the $O(n)$ model in some detail.
This loop model is related to the $A_2^{(2)}$ model \cite{Nienhuis90}.
Following Ref. \cite{Warnaar93} we use the
parametrization of the $A_2^{(2)}$ $R$-matrix in terms of $\tilde{u}$ and
$\tilde{\lambda}$, which is related to the parametrization used in Section
6 by $u=2 \I \tilde{u}$ and $\lambda= \I\tilde{\lambda} + \I \pi/2$. With a
suitable choice of gauge factors $a,b,c,d$ the bulk loop model weights are
given
by
\begin{eqnarray}
\rho_1 & = & \sin(3\tilde{\lambda}-\tilde{u})\sin(\tilde{u}) +
       \sin(2\tilde{\lambda})\sin(3\tilde{\lambda}),\nonumber\\
\rho_2 = \rho_3 & = &\epsilon_1 \sin(3\tilde{\lambda}-\tilde{u})
   \sin(2\tilde{\lambda}),\nonumber\\
\rho_4 = \rho_5 & = & \epsilon_2 \sin(\tilde{u})\sin(2\tilde{\lambda}),
    \nonumber\\
\rho_6 = \rho_7 & = & \sin(3\tilde{\lambda}-\tilde{u})\sin(\tilde{u}),
    \nonumber\\
\rho_8 &=& \sin(3\tilde{\lambda}-\tilde{u})
  \sin(2\tilde{\lambda}-\tilde{u}),\nonumber\\
\rho_9 &=& - \sin(\tilde{\lambda}-\tilde{u})\sin(\tilde{u}),
\label{eqn:lbulk}
\end{eqnarray}
where $\epsilon_1^2=\epsilon_2^2=1$ and $n=-2\cos(4\tilde{\lambda})$.
Referring back to equations (\ref{eqn:boa221}) to (\ref{eqn:boa223}) we see
that
for both choices of non quantum algebra invariant boundaries (cases (ii) and
(iii), but not case (i)) the constraints
(\ref{eqn:loop}) on the boundary weights are satisfied, the ratio being
$1 : e^{u}$ in both cases.
Thus, for both sets of integrable boundaries a loop model
interpretation is available.  For case (ii), with the boundary vertex weights
given by (\ref{eqn:boa222}), the corresponding boundary loop weights are given
by
\begin{eqnarray}
\rho_{10} = \rho_{12} & = & \sin[\smallfrac{1}{2}(3
\tilde{\lambda}-\tilde{u})],
\nonumber\\
\rho_{11} = \rho_{13} & = & \epsilon_1 \sin[\smallfrac{1}{2}(3 \tilde{\lambda}+
\tilde{u})],
\label{eqn:c2}
\end{eqnarray}
with the gauge factors used being the same ones as those which led to
(\ref{eqn:lbulk}).
As with vertex models, the bulk weights ($\rho_1,\ldots,\rho_9$),
left boundary weights ($\rho_{10},\rho_{11}$) and right boundary weights
($\rho_{12},\rho_{13}$) can be separately normalized without changing the
essential properties of the model. We have chosen one suitable normalization
here.

The relevant Bethe ansatz solution can be lifted from Section 6 and, up to
normalization factors, reads
\begin{equation}
\Lambda_D(\tilde{u}) =
\prod_{j=1}^M \frac{\sinh[\smallfrac{1}{2}(\I \tilde{u} + u_j) +
   \I \tilde{\lambda})]
             \sinh[\smallfrac{1}{2}(\I \tilde{u} - u_j) + \I \tilde{\lambda})]}
            {\sinh[\smallfrac{1}{2}(\I \tilde{u} + u_j) -  \I \tilde{\lambda})]
             \sinh[\smallfrac{1}{2}(\I \tilde{u} - u_j)  - \I
\tilde{\lambda})]}
\label{eqn:c2ev}
\end{equation}
where the numbers $u_j$ follow as roots of the equations ($1\leq k \leq M$)
\begin{eqnarray}
\lefteqn{
\left(\frac{\sinh[\smallfrac{1}{2}(u_k-\I \tilde{\lambda})]}
           {\sinh[\smallfrac{1}{2}(u_k+\I \tilde{\lambda})]}\right)^2
 \left(\frac{\sinh[\smallfrac{1}{2}(u_k - 2 \I \tilde{\lambda}-\I \tilde{u})]
           \sinh[\smallfrac{1}{2}(u_k -  2\I \tilde{\lambda} + \I \tilde{u})]}
           {\sinh[\smallfrac{1}{2}(u_k + 2 \I \tilde{\lambda}+ \I \tilde{u})]
           \sinh[\smallfrac{1}{2}(u_k +  2 \I \tilde{\lambda}-\I \tilde{u})]}
\right)^N} \hspace{90pt} \nonumber\\
&= & \prod_{j\neq k}^{M} \frac{\sinh[\smallfrac{1}{2}(u_k+u_j)-2 \I
    \tilde{\lambda}]
 \sinh[\smallfrac{1}{2}(u_k-u_j)-2\I \tilde{\lambda}]}
{\sinh[\smallfrac{1}{2}(u_k+u_j)+2\I \tilde{\lambda}]
 \sinh[\smallfrac{1}{2}(u_k-u_j)+2\I \tilde{\lambda}]} \nonumber\\
& & \times \; \frac{
\sinh[\smallfrac{1}{2}(u_k+u_j)+\I \tilde{\lambda}]
 \sinh[\smallfrac{1}{2}(u_k-u_j)+\I \tilde{\lambda}]}
{\sinh[\smallfrac{1}{2}(u_k+u_j)-\I \tilde{\lambda}]
 \sinh[\smallfrac{1}{2}(u_k-u_j)-\I \tilde{\lambda}]}.
\label{eqn:c2bae}
\end{eqnarray}
A special case of this solution has been derived using
the coordinate Bethe ansatz \cite{Batchelor93}. This pertains to
the case $\tilde{u} = \tilde{\lambda}$ which is equivalent to the $O(n)$ model
on the honeycomb lattice \cite{Nienhuis82,Nienhuis90,Reshetikhin91},
on account of the vanishing of
$\rho_9$ which enables all the other loop vertices to be ``pulled apart''
horizontally. The vertex weights given in Ref. \cite{Batchelor93} differ from
the ones obtained in Section 6 by a gauge transformation.

For case (iii), the boundary weights (\ref{eqn:c2}) are replaced by
\begin{eqnarray}
\rho_{10} = \rho_{12} & = & \cos[\smallfrac{1}{2}(3
\tilde{\lambda}-\tilde{u})],
\nonumber\\
\rho_{11} = \rho_{13} & = & \epsilon_1 \cos[\smallfrac{1}{2}(3 \tilde{\lambda}+
\tilde{u})].
\end{eqnarray}
The corresponding Bethe ansatz solution is given by (\ref{eqn:c2ev}) and
(\ref{eqn:c2bae}), with the sinh function in the squared term on the
left hand side of the latter replaced by cosh.
The honeycomb lattice Bethe ansatz solution corresponding to case (ii) with
$\tilde{u}=\tilde{\lambda}$ has been analysed for its critical
behaviour in Ref. \cite{Batchelor93} in connection with self-avoiding walks.
A study of the similarlyly interesting honeycomb
lattice solution corresponding to case (iii) will be presented elsewhere,
together with that for the square lattice solutions (general $u$).

\subsection{The $A_2^{(1)}$ loop model}

We now turn to the loop model connected with the $A_2^{(1)}$ vertex model
\cite{Reshetikhin91,Warnaar93}.
With a suitable choice of gauge factors, the bulk loop
weights are given by
\begin{eqnarray}
\rho_1  =  \rho_8 & = & \sinh(u+\lambda),\nonumber\\
\rho_2 = \rho_3 & = & \epsilon \sinh(\lambda), \nonumber\\
\rho_4 = \rho_5 & = & 0, \nonumber\\
\rho_6 = \rho_7 = \rho_9 & = & \sinh(u),
\end{eqnarray}
where $\epsilon^2=1$ and the loop fugacity is $n= -2\cosh(\lambda)$.
As seen from the $R$-matrix, the ratio $w_2 : w_3$ for this model is
$1 : e^{-2u}$.
An inspection of equations (\ref{eqn:boa211}) to (\ref{eqn:boa213})
shows that the criterion (\ref{eqn:loop})
is satisfied only for case (ii) and only if the parameters $\xi_{\pm}$ are
taken to $- \infty$. This represents the {\em only}
choice of integrable boundary
weights found in Section 5 for which the model has a loop model interpretation.
The boundary loop weights work out (using equations (\ref{eqn:necc})) to be
\begin{eqnarray}
\rho_{10} = \rho_{13} & = & e^{u/2},\nonumber\\
\rho_{11} = \rho_{12} & = & \epsilon e^{-u/2}.
\end{eqnarray}
The corresponding Bethe ansatz solution reads:
\begin{equation}
\Lambda_D(u)  =  \prod_{j=1}^{M_1} \frac{\sinh(\smallfrac{u}{2}
      + \mu_j^{(1)} - \smallfrac{\lambda}{2})
              \sinh(\smallfrac{u}{2}- \mu_j^{(1)} - \smallfrac{\lambda}{2})}
             {\sinh(\smallfrac{u}{2}+ \mu_j^{(1)} + \smallfrac{\lambda}{2})
              \sinh(\smallfrac{u}{2}- \mu_j^{(1)} + \smallfrac{\lambda}{2})},
\end{equation}
with the Bethe ansatz roots $\mu_k^{(1)}$ and $\mu_k^{(2)}$
satisfying the nested equations
\begin{eqnarray}
\lefteqn{
         \left[\frac{\sinh(\mu_k^{(1)}+\smallfrac{u}{2}+\smallfrac{\lambda}{2})
                    \sinh(\mu_k^{(1)}-\smallfrac{u}{2}+\smallfrac{\lambda}{2})}
                   {\sinh(\mu_k^{(1)}+\smallfrac{u}{2}-\smallfrac{\lambda}{2})
                    \sinh(\mu_k^{(1)}-\smallfrac{u}{2}
   -\smallfrac{\lambda}{2})}\right]^N}
 \hspace{20pt} \nonumber\\
& =&\prod_{j\neq k}^{M_1} \frac{\sinh(\mu_k^{(1)} + \mu_j^{(1)} + \lambda)
                    \sinh(\mu_k^{(1)} - \mu_j^{(1)} + \lambda)}
                   {\sinh(\mu_k^{(1)} + \mu_j^{(1)} - \lambda)
                    \sinh(\mu_k^{(1)} - \mu_j^{(1)} - \lambda)}\nonumber\\
& &  \times \; \prod_{j=1}^{M_2}\frac{\sinh(\mu_k^{(1)} + \mu_j^{(2)}-
                                 \smallfrac{\lambda}{2})
                    \sinh(\mu_k^{(1)} - \mu_j^{(2)}-\smallfrac{\lambda}{2})}
                   { \sinh(\mu_k^{(1)} + \mu_j^{(2)}+\smallfrac{\lambda}{2})
       \sinh(\mu_k^{(1)} - \mu_j^{(2)}+\smallfrac{\lambda}{2})},\nonumber\\
 1 & = & \prod_{j\neq k}^{M_2}  \frac{\sinh(\mu_k^{(2)} + \mu_j^{(2)} +
\lambda)
                    \sinh(\mu_k^{(2)} - \mu_j^{(2)} + \lambda)}
                   {\sinh(\mu_k^{(2)} + \mu_j^{(2)} - \lambda)
                    \sinh(\mu_k^{(2)} - \mu_j^{(2)} - \lambda)}\nonumber\\
 & &  \times \; \prod_{j=1}^{M_1}\frac{\sinh(\mu_k^{(2)} + \mu_j^{(1)}-
                                \smallfrac{\lambda}{2})
                    \sinh(\mu_k^{(2)} - \mu_j^{(1)}- \smallfrac{\lambda}{2})}
                    {\sinh(\mu_k^{(2)} + \mu_j^{(1)}+ \smallfrac{\lambda}{2})
                    \sinh(\mu_k^{(2)} - \mu_j^{(1)}+ \smallfrac{\lambda}{2})}.
\end{eqnarray}
The honeycomb limit $u = -\lambda$ of the $A_2^{(1)}$ model has been shown
\cite{Batchelor94} to yield the fully-packed loop model \cite{Blote94}
on the honeycomb lattice, which is equivalent to the zero-temperature limit
of the $O(n)$ model. For periodic boundary conditions the relevant Bethe
ansatz solution has been analysed \cite{Batchelor94} and applied to
the problem of Hamiltonian walks (i.e.\ self-avoiding walks which visit
every vertex) on the honeycomb lattice. The Bethe ansatz
solution obtained here for open boundary conditions is relevant to the
surface critical behaviour of Hamiltonian walks and will be discussed
elsewhere.

\subsection{The Zamolodchikov-Fateev loop model}

We will now investigate
the loop model connected with the Zamolodchikov-Fateev vertex model of
Section 4.
It turns out that a loop model interpretation is possible if and only if the
loops are allowed to intersect \cite{Nienhuis90}. To take this into account
an extra weight $\rho_0$ (see Fig. 11) is included in the definition of the
partition function $Z_{\rm loop}$. Furthermore, the loop fugacity is fixed
at $n= 2$ and the resulting model has an interpretation as the high-temperature
expansion of an $O(2)$ model on the square lattice \cite{Nienhuis90}.
%%%%%%%%%%%%%%%%%%  Figure 11 here  %%%%%%%%%%%%%%%%%%%%%
\begin{figure}[htb]
\epsfxsize = 1.7cm
\vbox{\vskip .8cm\hbox{\centerline{\epsffile{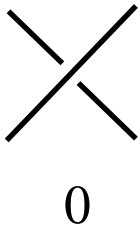}}}
%50%
\vskip .5cm \smallskip}
\caption{Extra vertex with Boltzmann weight $\rho_0$ needed if loops are
allowed to intersect}
\end{figure}
%%%%%%%%%%%%%%%%%%% End figure 11   %%%%%%%%%%%%%%%%%%%%%%%
The bulk loop model weights are then found to be
\begin{eqnarray}
\rho_1 & = & \sinh(u)\sinh(u+\lambda) + \sinh(\lambda)\sinh(2\lambda),
   \nonumber\\
\rho_2 = \rho_3 & = & \epsilon_1 \sinh(2\lambda)\sinh(u+\lambda),\nonumber\\
\rho_4 = \rho_5 & = & \epsilon_2 \sinh(u)\sinh(2\lambda),\nonumber\\
\rho_6 = \rho_7 & = & \sinh(u)\sinh(u+\lambda),\nonumber\\
\rho_8 & = & \cosh(u)\sinh(2\lambda)\sinh(u+\lambda),\nonumber\\
\rho_9 & = & -\sinh(u)\sinh(2u)\cosh(u+\lambda),\nonumber\\
\rho_0 & = & \sinh(u)\cosh(2\lambda)\sinh(u+\lambda),
\end{eqnarray}
where $\epsilon_1^2 = \epsilon_2^2 = 1$.
{}From the equations (\ref{eqn:bozf})
we see that the criterion (\ref{eqn:loop}) is satisfied
and thus a loop model interpretation is possible if and only if we set
$\xi_+ = \xi_- = \lambda/2$ in the integrable weights found in
Section 4.\footnote{Note that as with the $A_2^{(2)}$ and $A_2^{(1)}$ models,
the choice of $K$-matrices required for a loop model interpretation is {\em
not}
the choice which gives rise to quantum algebra-invariant spin chains.}
The boundary loop model weights are then found to be
\begin{eqnarray}
\rho_{10} = \rho_{12} & = & \sinh[\smallfrac{1}{2}(u+\lambda)],\nonumber\\
\rho_{11} = \rho_{13} & = & \epsilon_1 \sinh[\smallfrac{1}{2}
   (\lambda-u)].
\end{eqnarray}
The corresponding Bethe ansatz solution is given by
\begin{equation}
\Lambda_D(u) =
\prod_{j=1}^M \frac{\sinh(\smallfrac{u}{2}-u_j-\smallfrac{\lambda}{2})
             \sinh(\smallfrac{u}{2}+u_j-\smallfrac{\lambda}{2})}
             {\sinh(\smallfrac{u}{2}-u_j+\smallfrac{\lambda}{2})
             \sinh(\smallfrac{u}{2}+u_j+\smallfrac{\lambda}{2})},
\end{equation}
with the $u_j$ being roots of the equations
\begin{eqnarray}
\lefteqn{\left(\frac{\sinh(u_k -\frac{\lambda}{2})}
  {\sinh(u_k +\smallfrac{\lambda}{2})}\right)^2
  \left[\frac{\sinh(u_k-\smallfrac{u}{2}+\lambda)
    \sinh(u_k+\smallfrac{u}{2}+\lambda)}
   {\sinh(u_k-\smallfrac{u}{2}-\lambda)\sinh(u_k+\smallfrac{u}{2}-\lambda)}
  \right]^N}\hspace{120pt}\nonumber\\
\hspace{-80pt}
 & = & \prod_{j \neq k}^{M}\frac{\sinh(u_k-u_j+\lambda)\sinh(u_k+u_j+\lambda)}
    {\sinh(u_k-u_j-\lambda)\sinh(u_k+u_j-\lambda)}.
\end{eqnarray}

\subsection{Loop models related to the six-vertex model}
To conclude the study of loop models on the lattice ${\cal L}$, we now return
to the six-vertex model. The simplest loop model connected with the six-vertex
model is the so-called dense (i.e.\ all edges of ${\cal L}$ are covered) $O(n)$
model with partition function
\begin{equation}
Z_{\rm dense} = \sum_{{\cal G}} \rho_8^{m_8}\rho_9^{m_9}\rho_{10}^{m_{10}}
  \rho_{12}^{m_{12}} n^P,
\end{equation}
and Boltzmann weights
\begin{eqnarray}
\rho_8 & = & \sinh(u+\lambda) \nonumber\\
\rho_9 & = & \sinh(u).
\end{eqnarray}
The loop fugacity is
$n=-2\cosh(\lambda)$. Since the vertex weights $w_2$ and $w_3$ are zero, we
need to modify the criterion (\ref{eqn:loop}) which the boundary vertex
model weights need to satisfy in order to have a loop model interpretation.
It is easily seen that the new requirement is given by
\begin{equation}
w^L_3 : w^L_1 \;\;=\;\; w^R_1 : w^R_3
\;\; = \;\;w_{14} + s^{4} w_{16} : w_{19}.
\end{equation}
The required ratio is $1 : e^{\epsilon u}$ where $\epsilon^2=1$, and is
achieved only if the free parameters in the integrable six-vertex model
boundary weights found in Section 3 take on the values $\xi_{\pm} \rightarrow
 \infty$ or $\xi_{\pm} \rightarrow - \infty$.\footnote{Recall that
the choice $\xi_{\pm} \rightarrow - \infty$ leads
to a $U_q(su(2))$-invariant spin-$\smallfrac{1}{2}$ XXZ chain.}
The boundary weights can be conveniently normalized to
\begin{equation}
\rho_{10} = \rho_{12} = 1.
\end{equation}
The related Bethe ansatz solution is given by
\begin{eqnarray}
\Lambda_D(u) = \prod_{j=1}^M \frac{\sinh(\smallfrac{u}{2}-u_j -
   \smallfrac{\lambda}{2})
  \sinh(\smallfrac{u}{2}+u_i-\smallfrac{\lambda}{2} )}
  {\sinh(\smallfrac{u}{2}+u_i + \smallfrac{\lambda}{2})
   \sinh(\smallfrac{u}{2}-u_i+\smallfrac{\lambda}{2})}
\end{eqnarray}
with $u_j$ being roots of the equations
\begin{equation}
\left[\frac{\sinh(u_k + \smallfrac{u}{2} + \smallfrac{\lambda}{2})
    \sinh(u_k - \smallfrac{u}{2} + \smallfrac{\lambda}{2})}
  {\sinh(u_k + \smallfrac{u}{2} - \smallfrac{\lambda}{2})
    \sinh(u_k - \smallfrac{u}{2} - \smallfrac{\lambda}{2})}\right]^N
= \prod_{j\neq k}^M \frac{\sinh(u_k + u_j + \lambda)\sinh(u_k - u_j + \lambda)}
    {\sinh(u_k + u_j - \lambda)\sinh(u_k - u_j - \lambda)}.
\end{equation}

There is another loop model connected with the six-vertex model, a
dilute (i.e.\ not all edges of ${\cal L}$ are covered) $O(n)$ model.
This model has a partition function $Z_{\rm dilute}$ equivalent to
$Z_{\rm dense}$ but involves different loop weights. The equivalence can be
shown \cite{Warnaar92} by mapping $Z_{\rm dense}$ to  a 2-colour dense loop
model \cite{Warnaar93} with fugacities $n_1$ and $n_2$ such that $n_1+n_2=n$.
Setting $n_2=1$ and then summing over the second colour one arrives at
$Z_{\rm dilute}$ with weights
\begin{eqnarray}
\rho_1 & = & \sinh(u+\lambda) + \sinh(u) \nonumber\\
\rho_2 = \rho_3 = \rho_8 & = & \sinh(u+\lambda) \nonumber\\
\rho_4 = \rho_5  =  \rho_{9} & = & \sinh(u) \nonumber\\
\rho_{10} = \rho_{11} = \rho_{12} = \rho_{13} & = & 1.
\label{eqn:dilute}
\end{eqnarray}
The Bethe ansatz solution is the same as for the dense model, but the loop
fugacity $n$ is now given by $n=-2\cosh(\lambda)+1$. This solution has recently
been studied in \cite{Batchelor94b} in relation to walks on the Manhattan
lattice, both interacting self-avoiding walks at the
collapse temperature and Hamiltonian walks.

It is possible to perform the loop model to vertex model mapping directly
on the dilute $O(n)$-model to obtain a three-state Temperley-Lieb vertex
model (see e.g.\ Ref. \cite{Deguchi88})
with $R$-matrix $\check{R}(u)\equiv P R(u) =1 + f(u) \;U$ where
\begin{equation}
U_{ij}^{kl} = \delta_{i+j,N+1}\delta_{k+l,N+1} q^{-2S_i+k-2},
\end{equation}
$f(u)=\sinh(u)/\sinh(u+\lambda)$, $N=3$ and $q=e^{\lambda}$.
Integrable boundaries can then be derived along the lines of Section 2.
While it would be somewhat surprising, it is not inconceivable that
integrable boundaries other than those in (\ref{eqn:dilute}) could be
obtainable.

\section{Discussion}
\setcounter{equation}{0}

In this paper we have investigated integrable vertex models and loop models
on the lattice ${\cal L}$. We have applied Sklyanin's theory of $K$-matrices
and transfer
matrices $t(u,\boldomega)$ to derive integrable weights for various two- and
three-state vertex models and obtained Bethe ansatz solutions for their
eigenvalue spectra. We have also shown how some of these integrable vertex
models can be interpreted as loop models. It turns out that the choice of
$K$-matrices required for this does not usually coincide with the choice
which leads to quantum group-invariant spin chains.
Those $O(n)$ models amongst the integrable loop models
have particularly interesting physical applications (in the $n\rightarrow 0$
limit), being related to various polymer problems in two dimensions.
Another set of models of interest within the present framework are open
boundary versions of the restricted
solid-on-solid models constructed from loop models \cite{Warnaar92b}.
These include the dilute A-D-E models \cite{Warnaar92b,Roche92}
constructed from the $O(n)$ model of Section 7.1.  The study of the Bethe
ansatz solutions for the loop models found in this paper is currently under way
and will lead to exact results for various surface critical phenomena.

The method presented here can, of course, be applied to $N$-state models,
where $N>3$, and to related multi-colour loop models \cite{Warnaar93}.
However, there are still some interesting unsolved problems in the case of
two-state models. First of all, there is the problem of the elliptic
eight-vertex model on the lattice ${\cal L}$. $K$-matrices \cite{Cherednik84}
and commuting transfer matrices $t(u,\boldomega)$ \cite{Sklyanin88} have been
found for this model. Therefore, by the method of Section 2, integrable
boundary weights can be found. However, diagonalization of $t(u,\boldomega)$
is another matter altogether. Because the eight-vertex model does not have
the property of arrow-conservation across rows, the coordinate Bethe ansatz
fails. For the same reason, the algebraic Bethe ansatz presented in Section
3 -- so successful for
the six-vertex model  -- also fails, because there is no obvious
choice of pseudovacuum $|\Omega\rangle$. Baxter's ``commuting transfer
matrix method'' \cite{Baxter82} developed precisely for the eight-vertex
model with periodic boundaries\footnote{The generalized algebraic Bethe
ansatz used in Ref. \cite{Takhtadzhan79} uses a result from Ref.
\cite{Baxter82} to
obtain a family of pseuovacua.} might be able to be generalized to this
open boundary case.

More or less in the same category
is the also interesting problem of diagonalizing
the transfer matrix $t(u,\boldomega)$ for the six-vertex model with
{\em non-diagonal} $K$-matrices \cite{deVega93,Ghoshal93}. The corresponding
model on ${\cal L}$ has the feature of allowing sources or sinks of arrows
at the boundaries. Presumably some generalization of Baxter's commuting
transfer matrix method is also required to solve this problem.
This is presently under investigation.

Lastly, but by no means least,
integrable vertex models on the lattice ${\cal L}$
can be viewed as convenient regularizations of two dimensional quantum
field theories (the so-called ``light-cone approach'' -- for a recent
review, see Ref. \cite{deVega94c}). In particular, they are relevant to the
problem of calculating boundary $S$-matrices for various two dimensional
field theories with boundaries (see e.g.\ Refs.
\cite{Ghoshal93,Ghoshal93b,Fring93,Sasaki93,Corrigan94,Chim94,Grisaru94}).

\vspace{5pt}
\noindent
{\large \bf Acknowledgements}
\newline
We are grateful to Junji Suzuki  and Ole Warnaar for helpful discussions
on the coordinate Bethe ansatz and on $O(n)$ models.
This work has been supported by the Australian Research Council.

\end{document}